\documentclass[sigconf]{acmart}
\renewcommand\footnotetextcopyrightpermission[1]{} 
\setcopyright{none}

\usepackage[font={bf,sf,footnotesize}]{caption}
\usepackage[font={bf,sf,footnotesize}]{subfig}

\usepackage{graphicx}
\usepackage{booktabs}
\usepackage{epstopdf}
\usepackage{url}
\usepackage{placeins}
\usepackage{amsmath,amsfonts}
\usepackage{mathtools,mathrsfs}
\usepackage{multirow}
\usepackage{rotating}
\usepackage{pbox}
\usepackage[normalem]{ulem}

\usepackage{listings}

\newenvironment{sitemize}{
\begin{itemize}[noitemsep, parsep=0pt, partopsep=0pt, leftmargin=0.5em]
}{
\end{itemize}
}

\usepackage{cleveref}
\usepackage[utf8]{inputenc}
\crefname{section}{§}{§§}
\Crefname{section}{§}{§§}

\usepackage[10pt]{moresize}

\usepackage{xcolor}
\definecolor{darkgrey}{RGB}{70,70,70}
\definecolor{lightgrey}{RGB}{200,200,200}

\usepackage{enumitem}

\usepackage[linesnumbered,ruled,vlined]{algorithm2e}
\SetAlFnt{\scriptsize\tt}
\SetAlCapFnt{\scriptsize\sf}
\SetAlCapNameFnt{\scriptsize\sf}

\newcommand{\maciej}[1]{\textcolor{blue}{[Maciej: #1]}}
\newcommand{\htor}[1]{\textcolor{blue}{[Torsten: #1]}}

\newcommand{\goal}[1]{\noindent\textcolor{red}{[Goal: #1]}\par}

\newcommand{\nono}[1]{\textcolor{purple}{[Nono: #1]}}

\newcommand{\macb}[1]{\textbf{\textsf{#1}}}
\newcommand{\macbs}[1]{{\small\textbf{\textsf{#1}}}}

\usepackage[hang,flushmargin]{footmisc}

\usepackage{bm}

\usepackage{scalerel,stackengine}
\stackMath
\newcommand\rwh[1]{%
\savestack{\tmpbox}{\stretchto{%
  \scaleto{%
      \scalerel*[\widthof{\ensuremath{#1}}]{\kern-.6pt\bigwedge\kern-.6pt}%
          {\rule[-\textheight/2]{1ex}{\textheight}}
            }{\textheight}%
}{0.5ex}}%
\stackon[1pt]{#1}{\tmpbox}%
}

\def\HiLiGA{\leavevmode\rlap{\hbox to \hsize{\color{black!10}\leaders\hrule height 1\baselineskip depth 1ex\hfill}}}
\def\HiLiGB{\leavevmode\rlap{\hbox to \hsize{\color{black!25}\leaders\hrule height 1\baselineskip depth 1ex\hfill}}}
\def\HiLiGC{\leavevmode\rlap{\hbox to \hsize{\color{black!40}\leaders\hrule height 1\baselineskip depth 1ex\hfill}}}
\def\HiLiGD{\leavevmode\rlap{\hbox to \hsize{\color{black!55}\leaders\hrule height 1\baselineskip depth 1ex\hfill}}}
\def\HiLiGE{\leavevmode\rlap{\hbox to \hsize{\color{black!70}\leaders\hrule height 1\baselineskip depth 1ex\hfill}}}
\def\HiLiGF{\leavevmode\rlap{\hbox to \hsize{\color{black!85}\leaders\hrule height 1\baselineskip depth 1ex\hfill}}}

\usepackage{algpseudocode}
\usepackage{xcolor}
\makeatletter

\newcommand{\edgar}[1]{\textcolor{red}{[Edgar: #1]}}

\renewcommand{\goal}[1]{}
\renewcommand{\nono}[1]{}
\renewcommand{\maciej}[1]{}
\renewcommand{\edgar}[1]{}
\renewcommand{\htor}[1]{}

\newcommand{\macsubsection}[1]{\vspace{0em}\subsection{#1}\vspace{0em}}
\newcommand{\macsection}[1]{\vspace{0em}\section{#1}\vspace{0em}}

\newcommand{\acc}{\Leftarrow}
\newcommand{\krelax}[1]{#1-{\scshape relaxation}}
\newcommand{\kfilter}[1]{#1-{\scshape filter}}

\usepackage{tikz}
\usetikzlibrary{tikzmark}

\usepackage{xpatch}
\expandafter\xpatchcmd
   \csname pgfk@/tikz/every picture/.@cmd\endcsname
         {\thepage}{\arabic{page}}{}{}

\tikzstyle{comment} = [draw, fill=blue!70, text=white, text width=3cm, minimum height=1cm, rounded corners, align=left, font=\scriptsize]
\tikzstyle{background_alg} = [draw, fill=blue!20, opacity=0.4, inner sep=4pt, rounded corners=2pt]

\usetikzlibrary{shapes}
\usetikzlibrary{plotmarks}
\usetikzlibrary{calc,fit}

\newcommand\encircle[1]{%
  \tikz[baseline=(X.base)]
        \node (X) [draw, shape=circle, inner sep=0, fill=black, text=white, inner sep=-0.5pt] {\strut #1};}

        \newcommand\enbox[1]{%
          \tikz[baseline=(X.base)]
                \node (X) [draw, shape=regular polygon, regular polygon sides=4, inner sep=-1.5pt, fill=black, text=white] {\strut #1};}

\makeatletter
\global\let\tikz@ensure@dollar@catcode=\relax
\makeatother

\begin{document}

\title{To Push or To Pull: On Reducing Communication and Synchronization in Graph Computations}

\author{Maciej Besta$^1$, Michał Podstawski$^{2,3}$, Linus Groner$^1$, Edgar Solomonik$^4$, Torsten Hoefler$^1$}
       \affiliation{\vspace{0.3em}$^1$Department of Computer Science, ETH Zurich;\\ $^2$Perform Group Katowice; $^3$Katowice Institute of Information Technologies;\\ $^4$Department of Computer Science, University of Illinois at Urbana-Champaign\\
\vspace{0.3em}\small\sffamily maciej.besta@inf.ethz.ch, michal.podstawski@performgroup.com, gronerl@student.ethz.ch, solomon2@illinois.edu, htor@inf.ethz.ch\\
}

\begin{abstract}
We reduce the cost of communication and synchronization in graph processing
by analyzing the fastest way to process graphs:
pushing the updates to a shared state or pulling the updates to 
a private state.
We investigate the applicability of this push-pull dichotomy to various algorithms
and its impact on complexity, performance, and the amount of used locks, atomics, and reads/writes.
We consider 11 graph algorithms, 3 programming
models, 2 graph abstractions, and various families
of graphs.
The conducted analysis illustrates surprising differences between push and pull
variants of different algorithms in performance, speed of convergence, and code
complexity; the insights are backed up by performance data from
hardware counters. We use these findings to illustrate which variant is faster
for each algorithm and to develop generic strategies that enable even
higher speedups. 
Our insights can be used to accelerate graph processing engines or libraries on
both massively-parallel shared-memory machines as well as distributed-memory
systems.
\end{abstract}

\ccsdesc[100]{Computer systems organization~Parallel architectures}
\ccsdesc[100]{Computer systems organization~Multicore architectures}
\ccsdesc[100]{Computer systems organization~Distributed architectures}
\ccsdesc[500]{Theory of computation~Design and analysis of algorithms}
\ccsdesc[500]{Theory of computation~Graph algorithms analysis}
\ccsdesc[300]{Theory of computation~Shortest paths}
\ccsdesc[300]{Theory of computation~Parallel algorithms}
\ccsdesc[300]{Theory of computation~Shared memory algorithms}
\ccsdesc[300]{Theory of computation~Massively parallel algorithms}
\ccsdesc[300]{Computing methodologies~Parallel computing methodologies}
\ccsdesc[300]{Computing methodologies~Parallel algorithms}
\ccsdesc[300]{Computing methodologies~Shared memory algorithms}

\maketitle
\pagestyle{plain}

{\vspace{-0.5em}\noindent \textbf{This is an arXiv version of a paper published at\\ ACM HPDC'17 under the same title}}

{\vspace{1em}\small\noindent\textbf{Code:}\\\url{https://spcl.inf.ethz.ch/Research/Parallel\_Programming/PushPull}}

\macsection{INTRODUCTION}


\goal{Introduce the graph processing problems: synchronization and data movements}
Graph processing underlies many computational problems in social network
analysis, machine learning, computational science, and
others~\cite{DBLP:journals/ppl/LumsdaineGHB07}. Designing efficient
parallel graph algorithms is challenging due to several properties of
graph computations such as irregular communication patterns
or little locality~\cite{tate2014programming}. These properties lead to expensive synchronization
and movements of large data amounts on shared- and
distributed-memory (SM, DM) systems.

\goal{Introduce the PP notion in BFS}
Direction optimization in breadth-first search (BFS)~\cite{beamer2013direction}
is one of the mechanisms that are used to alleviate these issues. It
combines the traditional \emph{top-down} BFS (where vertices in the active
frontier iterate over all unvisited neighbors) with a \emph{bottom-up} scheme (where
unvisited vertices search for a neighboring vertex in the active frontier~\cite{suzumura2011performance}).
Combining these two approaches accelerates BFS by $\approx$2.4x on real-world graphs such as citation networks~\cite{beamer2013direction}.

\goal{Say that we generalize PP to many schemes}
We first illustrate that distinguishing between bottom-up and
top-down BFS can be generalized to many other graph algorithms, where
updates can be either \emph{pushed} by a thread to the shared
state (as in the top-down BFS), or \emph{pulled} to a thread's
private state (as in the bottom-up BFS).
As another example, consider a PageRank (PR) computation and assume a thread X
is responsible for a vertex $v$. X can either push $v$'s rank to update
$v$'s neighbors, or it can pull the ranks of $v$'s neighbors to update $v$~\cite{whang2015scalable}.
Despite many differences between PR and BFS (e.g., PR is
not a traversal), PR can similarly be viewed in the push-pull dichotomy.

This notion sparks various questions. Can pushing and pulling be applied
to \emph{any} graph algorithm? {How} to design push and pull variants of various algorithms?
Is pushing {or} pulling faster? {When} and {why}?  Does it depend on the
utilized programming model and abstraction? When and how can pushing
or pulling be accelerated? 

We seek to answer these and other questions and provide the first extensive 
analysis on the push-pull dichotomy in graph processing.
Now, this dichotomy was identified for some
algorithms~\cite{beamer2013direction, whang2015scalable} and was used in
several graph processing frameworks, such as Ligra~\cite{shun2013ligra} and
Gemini~\cite{zhu2016gemini}.
Yet, none of these works analyzes the differences in formulations,
complexity, and performance between the
two approaches for various algorithms, environments, or models. 

\begin{figure}[!h]
\centering
 \subfloat[Orkut network.]{
  \includegraphics[width=0.145\textwidth]{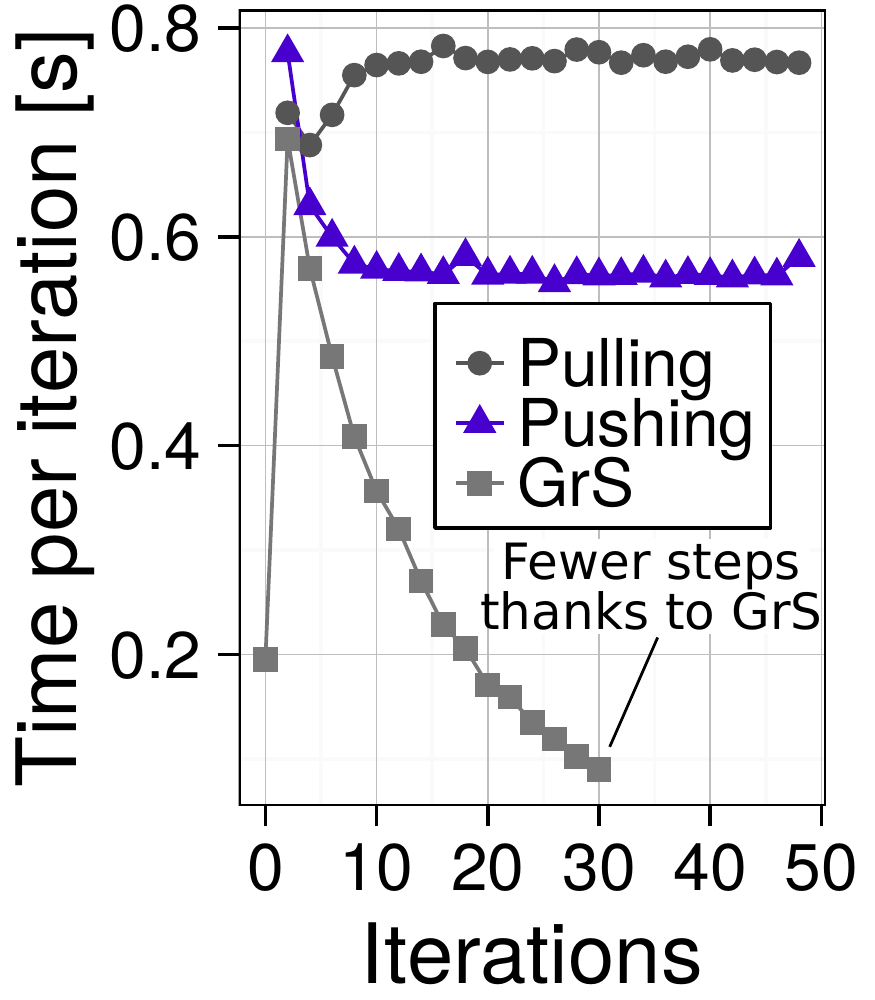}
  \label{fig:tc_orc}
 }
 \subfloat[Livejournal graph.]{
  \includegraphics[width=0.145\textwidth]{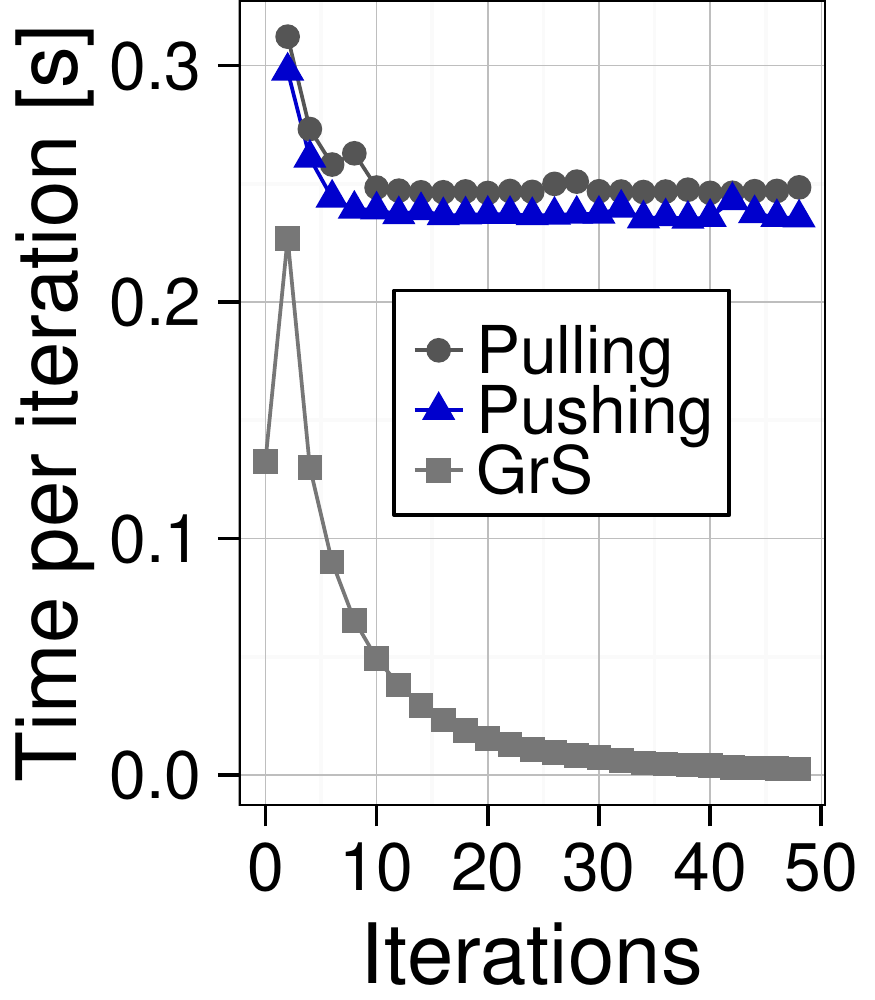}
  \label{fig:tc_ljn}
 } 
 \subfloat[CA road graph.]{
  \includegraphics[width=0.145\textwidth]{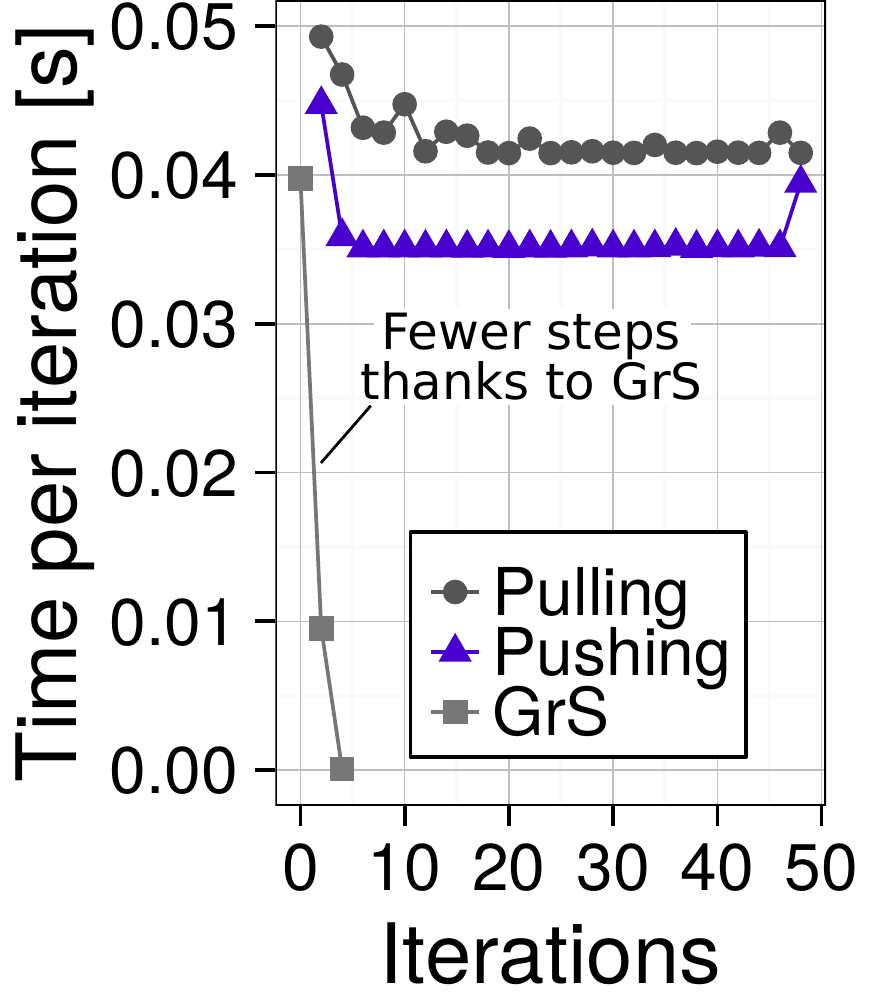}
  \label{fig:tc_orc}
 }
\caption{(\cref{sec:sm_perf}) Boman graph coloring~\cite{boman2005scalable} results and (\cref{sec:acc_perf}) the analysis of the strategy Greedy-Switch (GrS); single node of a Cray XC30, 16 threads.}
\label{fig:bgc_analysis}
\end{figure}

As a motivation, consider Figure~\ref{fig:bgc_analysis} with the results of our push/pull variants of graph coloring~\cite{boman2005scalable}.
They unveil consistent advantages of pushing. 
The figure also shows the speedup from
a strategy GrS (``Greedy-Switch'') that 
(1) reduces the number of memory access with a traversal-based graph coloring, and (2) switches between push- or pull-based scheme and an optimized
greedy variant.


We provide the following contributions:

\begin{sitemize}
\item We apply the push-pull dichotomy to various classes of graph algorithms
and obtain detailed formulations of centrality schemes, traversals,
calculating minimum spanning trees, graph coloring, and triangle counting.
We also show that several existing graph processing schemes
are included in the push-pull dichotomy.
\item We analyze pushing and pulling with PRAM 
and derive the differences in the amount of
synchronization and communication in both variants of the considered algorithms.
\item We analyze performance of push- and pull-based algorithms for
both SM and DM systems that represent fat-memory nodes and
supercomputers. Various programming models are incorporated, including
threading, Message Passing (MP), and Remote Memory Access
(RMA)~\cite{gerstenberger2013enabling} for various classes of graphs.
For detailed insights, we gather performance data (e.g., cache misses or issues braches and atomic instructions) using PAPI
counters.
\item We incorporate strategies to reduce the amount of
synchronization in pushing and memory accesses in pulling and illustrate that
they accelerate various algorithms. 
\item We provide performance insights that can be used to enhance
graph processing engines or libraries. 
\item Finally, we discuss whether the push-pull dichotomy is applicable
in the algebraic formulation of graph algorithms.
\end{sitemize}

\macsection{MODELS, NOTATION, CONCEPTS}

We first describe the necessary concepts. 

\macsubsection{Machine Model and Simulations}
\label{sec:background_models}

Parallel Random Access Machine (PRAM)~\cite{fortune1978parallelism} is a
well-known model of a parallel computer. There are $P$ processors that
exchange data by accessing cells of a shared memory of size $M$ cells.
They proceed in tightly-synchronized steps: no processor executes an instruction~$i+1$
before all processors complete an instruction~$i$. An instruction can be
a local computation or a read/write from/to the memory.
We use $S$ and $W$ to denote \emph{time} and \emph{work}: the longest execution path
and the total instruction count.
There are three PRAM variants with different rules for concurrent
memory accesses to the same cell. EREW prevents any concurrent accesses.
CREW allows for concurrent reads but only one write at a time. CRCW
enables any concurrent combination of reads/writes and 
it comes with multiple flavors that differently treat concurrent writes. We
use the Combining CRCW (CRCW-CB)~\cite{harris1994survey}: the value stored is an
associative and commutative combination of the written values.

Now, a simulation of one PRAM machine on another is a scheme that enables any
instruction from the former to be executed on the latter. Simulation schemes
are useful when one wants to port an algorithm developed for a stronger model
that is more convenient for designing algorithms (e.g., CRCW) to a weaker one
that models hardware more realistically (e.g., CREW). The used simulations are: 

\macbs{Simulating CRCW/CREW on CREW/EREW }
Any CRCW with $M$ cells can be simulated on an $MP$-cell CREW/EREW with a
slowdown of {\small$\Theta(\log n)$} and memory $MP$ (similarly to simulating a
CREW on an EREW)~\cite{harris1994survey}.

\macbs{Limiting $P$ (LP) }
A problem solvable on a $P$-processor PRAM in $S$
time can be solved on a $P'$-processor PRAM ($P' < P$) in time $S' =
\left\lceil\frac{S P}{P'}\right\rceil$ for a fixed memory size $M$.

\macsubsection{Graph Model, Layout, and Notation}


A tuple $(V,E)$ models an undirected graph $G$; $V$ is a set of vertices
and $E \subseteq V \times V$ is a set of edges;
$|V|=n$ and $|E|=m$. 
$d(v)$ and $N(v)$ are the degree and the neighbors of a vertex $v$.
The (non-negative) weight of an edge $(v,w)$ is {\small$\mathcal{W}_{(v,w)}$}.
We denote the maximum degrees for a given $G$ as $\hat{d}$,
$\hat{d}_{in}$ (in-degree), and $\hat{d}_{out}$ (out-degree).
The average degree is denoted with a
bar ($\overline{d}$). 
$G$'s diameter is $D$.

%
The neighbors of each $v$ form an
array. The arrays of all the vertices form a contiguous array
accessed by all the threads; we also store offsets into the array that
determine the beginning of the array of each vertex. The whole representation
takes $n+2m$ cells.

We partition $G$ by vertices (1D decomposition)~\cite{Catalyurek:2001:FHM:645609.663255}. 
We denote the number of used threads/processes as $P$.
We name a thread (process) that owns a given vertex $v$ as $t[v]$.
We focus on \emph{label-setting}
algorithms. 
In some of the considered schemes (e.g., PageRank) the number of
iterations $L$ is a user-specified parameter. 

\macsubsection{Atomic Operations}

Atomic operations (atomics) appear to the system as if they occur
instantaneously. They are used in lock-free graph computations to
perform fine-grained
updates~\cite{Gregor05theparallel,murphy2010introducing}. 
Here, we use CPU atomics that operate on integers.
We now present the relevant operations:

\noindent
\macb{Fetch-and-Add(*target, arg) (FAA)}: 
it increases \textsf{*target} by \textsf{arg} and
also returns \textsf{*target}'s previous value.

\noindent
\macb{Compare-and-Swap(*target, compare, value, *result) (CAS)}: if
\textsf{*target == compare} then \textsf{*target = value} and \textsf{*result =
true} are set, otherwise \textsf{*target} is not changed and \textsf{*result =
false}.

\macsubsection{Communication \& Synchronization}

Unless stated otherwise, we associate \emph{communication}
with: intra- or inter-node reads and writes, messages,
and collective operations other than barriers.
\emph{Synchronization} will indicate: any atomic operations, locks,
and any form of barrier synchronization.

\macsection{PUSH-PULL: APPLICABILITY}
\label{sec:applicability}


We first analyze what algorithms can be expressed in the push-pull (PP)
dichotomy; we revisit existing schemes and
discuss new cases. 

\macsubsection{PageRank (PR)} 

PR~\cite{Brin:1998:ALH:297805.297827} is an iterative centrality algorithm that obtains
the \emph{rank} of each vertex $v$: $r(v) = (1 - f)/|V| + \sum_{w \in N(v)} (f
\cdot r(w)/d(w))$; $f$ is the \emph{damp
factor}~\cite{Brin:1998:ALH:297805.297827}. PR 
is used to rank websites.

\macb{Pushing and Pulling?}
PR can be expressed in both~\cite{whang2015scalable}. In
the former, $t[v]$ updates all $v$'s neighbors with a value $r(v)/d(v)$ (it
pushes the value from $v$ to $N(v)$). In the latter, $t[v]$ updates $v$ with
values $r(u)/d(u)$, $u \in N(v)$ (it pulls the updates from $N(v)$ to $v$).

\macsubsection{Triangle Counting (TC)} 


In TC, one counts the number of triangles that each vertex $v \in V$ is a part
of; a triangle occurs if there exist edges $\{v,w\}, \{w,u\}, \{v,u\}$, where
$u,w \in V$ and $u,w \neq v, u \neq w$. TC is used in various statistics and
machine learning schemes~\cite{satish2014navigating} and libraries such as
igraph~\cite{csardi2006igraph}.

\macb{Pushing and Pulling?}
This algorithm is also expressible in both schemes. Consider a thread $t[v]$ that
counts the number of triangles associated with a vertex $v$ ($tc(v)$). It iterates
over $N(v)$ and, for each $u \in N(v)$, it iterates over $N(u)$ and checks if
$\exists w \in V, v \neq w \neq u$ such that $w \in N(u) \cap N(v)$;
the final sums are divided by 2 at the end.
If yes, then, in the push variant, it increments either one of
$tc(u)$ and $tc(w)$ while in the pull scheme it increments $tc(v)$.

\macsubsection{Breadth-First Search (BFS)} 

The goal of BFS~\cite{Cormen:2001:IA:580470} is to visit each vertex in $G$.
The algorithm starts with a specified \emph{root} vertex $r$ and visits
all its neighbors $N(r)$. Then, it visits all the unvisited neighbors of the root's
neighbors, and continues to process each level of neighbors in one step. 
BFS represents graph traversals and is used 
the HPC benchmark Graph500~\cite{murphy2010introducing}.

\macb{Pushing and Pulling?}
There exist both variants. The former is the traditional
\emph{top-down} BFS where $t[v]$ (if $v$ is in a frontier) checks each unvisited
$u \in N(v)$ and adds it to the next frontier $F$ (it pushes the updates from $v$ to
$N(v)$). The latter is the \emph{bottom-up}
approach~\cite{beamer2013direction,suzumura2011performance}: in each
iteration every unvisited vertex $u$ is tested if it has a parent in $F$ 
(the updates are pulled from $N(u)$ to $u$).

\macsubsection{Single Source Shortest Path (SSSP)} 

SSSP outputs the distance from a selected source vertex $s$ to all other
vertices. We consider $\Delta$-Stepping (SSSP-$\Delta$)~\cite{meyer2003delta}
that combines the well-known Dijkstra's and Bellman-Ford algorithms by
trading work-optimality for more parallelism. It groups vertices into
\emph{buckets} and only vertices in one bucket can be processed in
parallel. 
SSSP has applications in, e.g., operations research.  

\macb{Pushing and Pulling?}
Both are applicable when relaxing edges of each vertex $v$ from
the current bucket. In the former, $v$ {pushes} relaxation requests to
its neighbors in the buckets with unsettled vertices. In the latter,
vertices in unsettled buckets look for their neighbors in the current bucket and
perform ({pull}) relaxations.
A similar scheme was used in the DM implementation of SSSP-$\Delta$~\cite{chakaravarthy2014scalable}.

\macsubsection{Betweenness Centrality (BC)} 


BC measures the importance of a vertex $v$ based on the number of
shortest paths that lead through $v$. Let $\sigma_{st}$ be the number of shortest
paths between two vertices $s,t$, and let $\sigma_{st}(v)$ be the
number of such paths that lead through $v$. BC of $v$ equals $bc(v) =
\sum_{s \neq v \neq t \in V} \frac{\sigma_{st}(v)}{\sigma_{st}}$.
Here, we consider Brandes'
algorithm~\cite{prountzos2013betweenness, brandes2001faster, solomonik2017scaling}. Define the
dependency of a source vertex $s$ on $v$ as: $\delta_{s}(v) = \sum_{t \in V}
\frac{\sigma_{st}(v)}{\sigma_{st}}$. Then, we have $bc(v) = \sum_{s \neq v \in
V} \delta_{s}(v)$ where $\delta_{s}(v)$ satisfies the following recurrence:
$\delta_{s}(v) = \sum_{w: v \in pred(s,w)} \frac{\sigma_{sv}}{\sigma_{sw}} (1 +
\delta_{s}(w))$; $pred(s,w)$ is a list of immediate \emph{predecessors} of $w$ in the
shortest paths from $s$ to $w$.
Brandes' scheme uses this recurrence to compute $bc(v)$ in two phases.  First,
BFS or SSSP traversals compute $pred(s,v)$ and $\sigma_{sv}$,
$\forall_{s,v \in V}$, obtaining a tree $\mathcal{T}$ over $G$. Next, $\mathcal{T}$ is traversed
backwards (from the highest to the lowest distance) to compute $\delta_{s}(v)$
and $bc(v)$ based on the equations above.
BC is a complex centrality scheme used in
biology, transportation, and terrorism prevention~\cite{bader2007approximating}.

\macb{Pushing and Pulling?}
Both parts of Brandes BC can be expressed using push and pull. 
The first phase can compute shortest path information using either
top-down or bottom-up BFS or push- and pull-based versions of SSSP. 
The second phase (backward accumulation) may also be cast as BFS from a starting frontier.
In particular, one can either push partial centrality scores to predecessors or pull them from 
lists of \emph{successors}~\cite{madduri2009faster}.

\macsubsection{Graph Coloring (GC)} 

GC assigns colors to vertices so that no two incident
vertices share the same color and the number of colors is minimized. We
consider Boman graph coloring (BGC)~\cite{boman2005scalable}.
Here, each iteration has two phases. In phase~1, colors are assigned to
vertices owned by each thread (i.e., to each partition $\mathcal{P} \in
\mathscr{P}$) separately
without considering other partitions ($\mathscr{P}$ denotes a set of all partitions). 
The maximum number of available colors can be specified as a parameter $\mathcal{C}$.
In phase~2, \emph{border} vertices (i.e.,
vertices with at least one edge leading to another partition; they form a set
$\mathcal{B}$) are verified for conflicts. If there are any, the
colors are reassigned. This may cause conflicts within partitions, which are
resolved during the next iteration. More iterations $L$ may 
improve a solution (fewer colors used).
GC has multiple applications in scheduling and pattern matching.  

\macb{Pushing and Pulling?}
Both can be used in phase~2.
For every border vertex
$v$, each $u \in N(v)$ ($t[u] \neq t[v]$) is analyzed. If $v$ and $u$ share the
assigned color, then either $u$'s or $v$'s color is scheduled for a
change (the update is pushed to or pulled from $N(v)$).

\macsubsection{Minimum Spanning Tree (MST)} 

The goal of MST is to derive a spanning tree of $G$ with the lowest sum of the
included edge weights. 
The classical sequential algorithms: Prim~\cite{Cormen:2001:IA:580470} and Kruskal~\cite{Cormen:2001:IA:580470} lack parallelism.
Therefore, we focus on the Boruvka~\cite{boruuvka1926jistem} algorithm (more details on pushing and pulling in Prim and Kruskal are still provided in the technical report).
In Boruvka, each vertex is first associated with its own supervertex. In each
iteration, two incident supervertices are merged into one along an edge $e_m$ of a
minimum weight. The algorithm proceeds until there is only one supervertex
left. The selected minimum edges form the MST.
MST algorithms are utilized in problems such as the design of broadcast
trees~\cite{Cormen:2001:IA:580470}.

\macb{Pushing and Pulling in Boruvka?}
First, selecting $e_m$ adjacent to a given supervertex can be
done by pushing (each supervertex overrides adjacent supervertices
and their tentative minimal edges if it has a less expensive one) or by pulling (each
supervertex picks its own $e_m$).
Next, merging adjacent supervertices can also be done with pushing or pulling.
Assume that each thread owns a number of supervertices. Now, it can
either push the changes to the supervertices owned by other threads, or pull
the information on the adjacent supervertices and only modify its owned ones.

\macsubsection{Push-Pull Insights}


First, we present a generic difference between pushing and pulling. Recall that
$t[v]$ indicates the thread that owns $v$. Define $t\leadsto v$ to be
true if $t$ modifies $v$ during the execution of a given algorithm ($t \leadsto v \Leftrightarrow t$ modifies $v$). Then

\begin{gather} 
\left(\text{Algorithm uses pushing}\right) \Leftrightarrow \left(\exists_{t \in \{1..T\}, v \in V}\ t \leadsto v \land t \neq t[v] \right) \nonumber\\
\left(\text{Algorithm uses pulling}\right) \Leftrightarrow \left(\forall_{t \in \{1..T\}, v \in V}\ t \leadsto v \Rightarrow t = t[v] \right) \nonumber
\end{gather}
In pushing, any thread $t$ may access and modify any vertex $v \in V$ so that
we may have $t \neq t[v]$. In pulling, $t$ can only modify its assigned
vertices: $t[v] = t$ for any $v$ modified by $t$. 
In \cref{sec:th}, we show that this property determines that pulling requires
less synchronization compared to pushing.
However, pushing can often be done with less work, when only a subset 
of vertices needs to update its neighbors.

Second, our analysis shows that the push-pull dichotomy can be used in 
two algorithm classes:
\emph{iterative} schemes (PR, TC, GC, Boruvka MST) that derive some vertex properties and perhaps
proceed in iterations until some convergence condition is met, and
\emph{traversals} (BFS, SSSP-$\Delta$, BC).

\macsection{THEORETICAL ANALYSIS}
\label{sec:th}

We now derive detailed specifications of push and pull algorithm variants and
use them to investigate the differences between pushing and pulling.
We (1) identify \emph{read and write conflicts},
(2) conduct complexity analyses, and (3) investigate the 
amount of required atomics or locks.
We focus on the CRCW-CB and CREW models.
There exist past works on the parallel complexity of the considered algorithms~\cite{goel2012complexity, leiserson2010work, meyer2003delta, boman2005scalable, madduri2009faster, prountzos2013betweenness, bader2004fast}.
Yet, we are the first to investigate the differences between pushing and pulling
variants.
 
\macb{Algorithm Listings}
Our schemes have multiple variants as many nested loops can be
parallel; we indicate them with \texttt{[in par]}. 
Unless specified otherwise, we only consider the loops without square brackets in complexity analyses.
We mark the read/write conflicts in the parts of the code related to pushing or pulling with
$\text{\small\encircle{R}}$/$\text{\small\encircle{W}}$,
respectively.
We indicate the data type in the modified memory cell to be either integer
($\text{\scriptsize\enbox{i\bf}}$) or float ($\text{\scriptsize\enbox{f\bf}}$).
\begin{tikzpicture}[remember picture,overlay,pin distance=0cm]
\draw[fill=black!15, opacity=1, inner sep=4pt, rounded corners=4pt]
  ([shift={(-0.25em,0.8em)}]pic cs:explain-grey) 
    rectangle
  ([shift={(7.5em,-0.4em)}]pic cs:explain-grey);
\end{tikzpicture}
Finally, we use \tikzmark{explain-grey}grey backgrounds to indicate
pushing/pulling variants.


\macb{Cost Derivations}
We consider up to one processor per vertex, $P\leq n$ (and $P > \hat{d}$). 
Thus, pulling avoids write-conflicts, 
as each thread accumulates updates for a given vertex.
Still, pushing can update the same vertices multiple times at every iteration.

We formulate cost analyses of all algorithms via the primitives \krelax{$k$} and \kfilter{$k$}. 
\krelax{$k$} corresponds to simultaneously propagating updates from/to $k$ vertices to/from one of their neighbors for pushing/pulling.
\kfilter{$k$} is used to extract the vertices updated in one or more \krelax{$k$}s, and is non-trivial only when pushing updates.
We let $\bar{k}=\max(1,k/P)$ and quantify the cost of these primitives.
When pulling, \krelax{$k$} takes $\mathcal O(\bar{k})$ time and $\mathcal O(k)$ work.
A \kfilter{$k$} invocation requires $O(\log(P)+\bar{k})$ time and $O(\min(k,n))$ work via a prefix sum.

When pushing, the cost of \krelax{$k$} depends on the PRAM model. 
In the CRCW-CB model, \krelax{$k$} takes $\mathcal O(\bar{k})$ time and $\mathcal O(k)$ work.
In the CREW model, \krelax{$k$} can be processed in $\mathcal O(\bar{k}\log(\hat{d}))$ time via binary-tree reductions.
To update each vertex of degree $d$ in the CREW model, we use a binary merge-tree with $d$ leaves.
Over all trees, at most $k$ of $m$ leaves contain actual updates.
We can avoid work for all nodes that are the roots of subtrees that do not contain updates, effectively computing
 a forest of incomplete binary trees with a total of $k$ leaves and maximum height $\mathcal O(\log(\hat{d}))$.
Each of $P$ processors propagates $k/P$ updates up the complete binary merge-tree associated with its vertices (requiring no setup time) in $\mathcal O(\bar{k}\log(\hat{d}))$ time with a total of $\mathcal O(k\log(\hat{d}))$ work.



\begin{tikzpicture}[remember picture,overlay,pin distance=0cm]
\draw[fill=black!15, opacity=1, inner sep=4pt, rounded corners=4pt]
  ([shift={(-0.25em,0.55em)}]pic cs:pr-ps) 
    rectangle
  ([shift={(25em,-0.6em)}]pic cs:pr-ps);
\fill ([shift={(23.25em,0.0em)}]pic cs:pr-ps) node[inner sep=2pt, rounded corners, text=white, fill=black!75, font=\tiny] {\macb{PUSHING}};
\draw[fill=black!15, opacity=1, inner sep=4pt, rounded corners=4pt]
  ([shift={(-0.25em,-0.6em)}]pic cs:pr-ps) 
    rectangle
  ([shift={(25em,-1.75em)}]pic cs:pr-ps);
\fill ([shift={(23.25em,-1.2em)}]pic cs:pr-ps) node[inner sep=2pt, thin,rounded corners, text=white, fill=black!75, font=\tiny] {\textsf{\textbf{PULLING}}};
\end{tikzpicture}

\begin{lstlisting}[aboveskip=0em,float=!h,label=lst:pram_pr,caption=(\cref{sec:pr_pram}) Push- and pull-based PageRank.]
/* Input: a graph $G$, a number of steps $L$, the damp parameter $f$
  Output: An array of ranks pr[1..$n$] */

function PR($G$,$L$,$f$) {
  pr[1..$v$] = [$f$..$f$]; //Initialize PR values.
  for($l=1$; $l < L$; ++$l$) { 
    new_pr[1..$n$] = [0..0]; 
    for $v \in V$ do in par { |\label{ln:pr-inner-start}|
      update_pr(); new_pr[$v$] += $(1-f)/n$; pr[$v$] = new_pr[$v$];
} } }|\label{ln:pr-inner-end}|

function update_pr() {
  for $u \in N(v)$ do [in par] {
|\tikzmark{pr-ps}|    {new_pr[$u$] += $(f\cdot$pr[$v$]$)/d(v)$ $\text{\encircle{W}}$ $\text{\enbox{f\bf}}$;}|\label{ln:write-con}|

|\tikzmark{pr-pl}|    {new_pr[$v$] += $(f\cdot$pr[$u$]$)/d(u)$ $\text{\encircle{R}}$;}|\label{ln:read-con}|
} }
\end{lstlisting}

\macsubsection{PageRank}
\label{sec:pr_pram}

PR (Algorithm~\ref{lst:pram_pr}) performs $\mathcal O(L)$ steps of power iteration.
For each step of power iteration, \krelax{$k_i$} is called for $i\in\{1,\ldots, \hat{d}\}$ with $\sum_{i=1}^{\hat{d}} = m$.
Thus the PRAM complexities of PR are
(1) $\mathcal O(L(m/P+\hat{d}))$ time and $\mathcal O(Lm)$ work using pulling,
(2) $\mathcal O(L(m/P+\hat{d}))$ time and $\mathcal O(Lm)$ work in pushing in CRCW-CB,
and (3) $\mathcal O(L\log(\hat{d})(m/P +\hat{d}))$ time and $\mathcal O(Lm\log(\hat{d}))$ work using pushing in CREW.

\macb{Conflicts}
Pushing/pulling entail {\small$\mathcal{O}(L m)$} write/read conflicts.


\macb{Atomics/Locks}
Pulling does not require any such operations.
Contrarily, pushing comes with write conflicts to
floats. To the best of our knowledge, no CPUs offer atomics
operating on such values. Thus, {\small$\mathcal{O}(L m)$} locks are issued.
%

\macsubsection{Triangle Counting}
\label{sec:pram_tc}

TC is shown in Algorithm~\ref{lst:pram_tc}; this is a simple parallelization of
the well-known NodeIterator scheme~\cite{schank2007algorithmic}.
It employs \krelax{$k_i$} for $i\in\{1,\ldots, \hat{d}^2\}$ with $\sum_{i=1}^{\hat{d}} = O(m\hat{d})$.
Thus the PRAM complexities of TC are
(1) $\mathcal{O}(\hat{d}(m/P+\hat{d}))$ time and $\mathcal O(m\hat{d})$ work using pulling,
(2) $\mathcal{O}(\hat{d}(m/P+\hat{d}))$ time and $\mathcal O(m\hat{d})$ work using pushing in CRCW-CB, 
and (3) $\mathcal{O}(\hat{d}\log(\hat{d})(m/P+\hat{d}))$ time and $\mathcal O(m\hat{d}\log(\hat{d}))$ work 
using pushing in CREW.
One can leverage more than $n$ processors to lower the PRAM time-complexity of TC~\cite{7113280}.


\macb{Conflicts}
Both variants generate $\mathcal{O}(m \hat{d})$ read conflicts;
pushing also has $\mathcal{O}(m \hat{d})$ write conflicts.

\macb{Atomics/Locks}
We use FAA atomics to resolve write conflicts.

\begin{tikzpicture}[remember picture,overlay,pin distance=0cm]
\draw[fill=black!15, opacity=1, inner sep=4pt, rounded corners=4pt]
  ([shift={(-0.25em,0.55em)}]pic cs:tc-ps) 
    rectangle
  ([shift={(25em,-0.6em)}]pic cs:tc-ps);
\fill ([shift={(23.25em,0em)}]pic cs:tc-ps) node[inner sep=2pt, rounded corners  , text=white, fill=black!75, font=\tiny] {\macb{PUSHING}};
\draw[fill=black!15, opacity=1, inner sep=4pt, rounded corners=4pt]
  ([shift={(-0.25em,-0.6em)}]pic cs:tc-ps) 
    rectangle
  ([shift={(25em,-1.75em)}]pic cs:tc-ps);
\fill ([shift={(23.25em,-1.2em)}]pic cs:tc-ps) node[inner sep=2pt, thin,rounded corners, text=white, fill=black!75, font=\tiny] {\textsf{\textbf{PULLING}}};
\end{tikzpicture}

\begin{lstlisting}[aboveskip=0em,belowskip=-1em,float=!h,label=lst:pram_tc,caption=(\cref{sec:pram_tc}) Push- and pull-based Triangle Counting.]
/* Input: a graph $G$. Output: An array of triangle counts 
 * tc[1..$n$] that each vertex belongs to. */

function TC($G$) {tc[1..$n$] = [$0$..$0$]
  for $v \in V$ do in par 
    for $w_1 \in N(v)$ do [in par] 
      for $w_2 \in N(v)$ do [in par] 
        if adj($w_1$,$w_2$) $\text{\encircle{R}}$ update_tc();   
  tc[1..$n$] = [tc[1]/2 .. tc[$n$]/2]; }
function update_tc() {
|\tikzmark{tc-ps}|  {++tc[$w_1$]; /* or ++tc[$w_2$]. */} $\text{\encircle{W}}$ $\text{\enbox{i\bf}}$ 

|\tikzmark{tc-pl}|  {++tc[$v$];} 
} 
\end{lstlisting}

\macsubsection{Breadth-First Search}
\label{sec:pram_bfs}


BFS is shown in Algorithm~\ref{lst:pram_bfs}.
We define a generalized version of BFS, where vertices enter the frontier only after a given number of neighbors have been in the frontier.
The standard BFS is obtained by setting this number to $1$, but to use BFS from within BC, we will employ a counter specific to each vertex.
The BFS pseudo-code also employs a given accumulation operator to compute values for each vertex as a function of values of its predecessors in the BFS tree.
Our analysis assumes this operator is commutative and associative.
The frontier \texttt{F} is represented as a single array while \texttt{my\_F} is private
for each process and contains vertices explored at each iteration. 
All \texttt{my\_F}s are repeatedly merged into the next \texttt{F} (Line~\ref{ln:bfs-merge}).
We let $f_i$ be the size of \texttt{F} in the $i$th iteration of the while loop.


\begin{tikzpicture}[remember picture,overlay,pin distance=0cm]
\draw[fill=black!10, opacity=1, inner sep=4pt, rounded corners=4pt]
  ([shift={(-0.25em,0.55em)}]pic cs:bfs-ps) 
    rectangle
  ([shift={(25em,-5.2em)}]pic cs:bfs-ps);
\fill ([shift={(23em,-0.2em)}]pic cs:bfs-ps) node[rounded corners, text=white, fill=black!75, font=\tiny] {\macb{PUSHING}};
\draw[fill=black!10, opacity=1, inner sep=4pt, rounded corners=4pt]
  ([shift={(-0.25em,-5.2em)}]pic cs:bfs-ps) 
    rectangle
  ([shift={(25em,-11.2em)}]pic cs:bfs-ps);
\fill ([shift={(23em,-6.0em)}]pic cs:bfs-ps) node[thin,rounded corners, text=white, fill=black!75, font=\tiny] {\macb{PULLING}};
\end{tikzpicture}

\begin{lstlisting}[aboveskip=0em, float=!h,label=lst:pram_bfs,caption=(\cref{sec:pram_bfs}) Push- and pull-based Breadth-First Search.]
/* Input: a graph $G$, a set of ready counters and initial values R0 for each node, and an accumulation operator $\acc$.
 * Output: R[1..$n$] where R[F[$i$]]=R0[$i$] and other otherwise contains accumulation of all R values of predecessors. */

function BFS($G$,ready,R0,$\acc$) {
  my_F[1..$P$] = [$\emptyset$..$\emptyset$]; R = R0; F$\subset V$, such that for each $v\in$F, ready[$v$]=0;
  while (F $\neq \emptyset$)
    explore_my_F(); { |\label{ln:pr-inner-start}|
    F = my_F[1] $\cup$ my_F[2] $\cup$ .. $\cup$ my_F[$P$]; |\label{ln:bfs-merge}| } }

function explore_my_F() { 
|\tikzmark{bfs-ps}| for $v \in$ F do in par 
    for $w \in N(v)$ do [in par] 
      if ready[$w$] > 0 $\text{\encircle{R}}$ 
        R[$w$] $\acc$ R[$v$] $\text{\encircle{W}}$; 
    for $w \in N(v)$ do [in par] {
      ready[$w$]--; 
      if ready[$w$]==0 { my_F[$p_{ID}$] = my_F[$p_{ID}$] $ \cup $ $\{w\}$; } }
      
|\tikzmark{bfs-pl}| for $v \in V$ do in par { |\label{ln:pr-inner-start}|
    if ready[$v$] > 0 {
      for $w \in N(v)$ do [in par] {
        if $w \in$ F $\text{\encircle{R}}$ {
          R[$v$] $\acc$ R[$w$];
          ready[$v$]--; 
          if ready[$v$] == 0 { my_F[$p_{ID}$] = my_F[$p_{ID}$] $ \cup $ $\{v\}$; }
} } } }
\end{lstlisting}

The call to \texttt{explore\_my\_F} in pulling requires checking all edges, so it takes $\mathcal O(m/P+\hat{d})$ time and $\mathcal O(m)$ work.
The call to \texttt{explore\_my\_F} in pushing needs $\mathcal O(\hat{d})$ consecutive \krelax{$f_i$}s, so it takes $\mathcal O(\bar{f}_i\hat{d})$ time where $\bar{f}_i=\max(1,f_i/P)$ and work $\mathcal O(f_i\hat{d})$ in CRCW-CB (and $\mathcal O(\log(\hat{d}))$ more in CREW).
Second, the merge of frontiers can be done via a \krelax{$\hat{d}f_i$} and, in pushing, a \kfilter{$\hat{d}f_i$}.
The \kfilter{$\hat{d}f_i$} is not required in pulling, since we check whether each vertex is in the frontier anyway.
In pushing, the merge requires $\mathcal O(\log(P)+\hat{d}f_i/P)$ time and $\mathcal O(\min(\hat{d}f_i,n))$ work.

Thus, for a graph of diameter $D$ (with $D$ while-loop iterations) we derive the total cost using the fact that $\sum_{i=1}^D f_i= n$, obtaining:
(1) $\mathcal O(D(m/P+\hat{d}))$ time and $\mathcal O(Dm)$ work in pulling,
(2) $\mathcal O(m/P + D(\hat{d}+\log(P)))$ time and $\mathcal O(m)$ work in pushing in CRCW,
and (3) a factor of $O(\log(\hat{d}))$ more time and work 
in the CREW model.
It is possible to achieve a lower time-complexity for BFS, especially if willing to sacrifice work-efficiency~\cite{gazit1988improved}.

\macb{Conflicts}
There are $\mathcal O(m)$ write conflicts in pushing; pulling involves $\mathcal O(Dm)$ read conflicts.


\macb{Atomics/Locks}
Pushing requires $\mathcal{O}(m)$ CAS atomics.


\macsubsection{{\large$\Delta$}-Stepping SSSP}
\label{sec:pram_ssspd}

\begin{tikzpicture}[remember picture,overlay,pin distance=0cm]
\draw[fill=black!10, opacity=1, inner sep=4pt, rounded corners=4pt]
  ([shift={(-0.25em,0.55em)}]pic cs:sssp-ps) 
    rectangle
  ([shift={(25em,-6.6em)}]pic cs:sssp-ps);
\fill ([shift={(23em,-0.2em)}]pic cs:sssp-ps) node[rounded corners, text=white, fill=black!75, font=\tiny] {\macb{PUSHING}};
\draw[fill=black!10, opacity=1, inner sep=4pt, rounded corners=4pt]
  ([shift={(-0.25em,-6.6em)}]pic cs:sssp-ps) 
    rectangle
  ([shift={(25em,-12.9em)}]pic cs:sssp-ps);
\fill ([shift={(23em,-7.2em)}]pic cs:sssp-ps) node[thin,rounded corners, text=white, fill=black!75, font=\tiny] {\macb{PULLING}};
\end{tikzpicture}

\begin{lstlisting}[aboveskip=-2em, belowskip=-2em, float=h!, label=lst:pram_ssspd,caption=(\cref{sec:pram_ssspd}) Push- and pull-based $\Delta$-Stepping SSSP.]
/* Input: a graph $G$, a vertex $r$, the $\Delta$ parameter. 
   Output: An array of distances d */

function $\Delta$-Stepping($G$, $r$, $\Delta$){
 bckt=[$\infty$..$\infty$]; d=[$\infty$..$\infty$]; active=[false..false]; 
 bckt_set={0}; bckt[$r$]=0; d[$r$]=0; active[$r$]=true; itr=0;

 for $b \in$ bckt_set do { //For every bucket do...
  do {bckt_empty = false; //Process $b$ until it is empty.
  process_buckets();} while(!bckt_empty); } }

function process_buckets() { 
|\tikzmark{sssp-ps}| for $v \in$ bckt_set[b] do in par
  if(bckt[v]==b && (itr == 0 or active[$v$])) {
   active[$v$] = false; //Now, expand $v$'s neighbors.
   for $w \in N(v)$ {weight = d[$v$] + $\mathcal{W}_{(v,w)}$;
    if(weight < d[$w$]) { $\text{\encircle{R}}$ //Proceed to relax $w$.
     new_b = weight/$\Delta$; bckt[$v$] = new_b;
     bckt_set[new_b] = bckt_set[new_b] $\cup$ {$w$};}
    d[$w$] = weight;	$\text{\encircle{W}}$ $\text{\enbox{i\bf}}$;
    if(bckt[$w$]==b)$\text{\encircle{R}}$ {active[$w$]=true; bckt_empty=true;}}} $\text{\encircle{R}}$
|\tikzmark{sssp-pl}| for $v \in V$ do in par 
  if(d[$v$] > b) {for $w \in N(v)$ do {
    if(bckt[$w$] == b && (active[$w$] or itr == 0)) {$\text{\encircle{R}}$
     weight = d[$w$] + $\mathcal{W}_{(w,v)}$	$\text{\encircle{R}}$;
     if(weight < d[$v$]) {d[$v$]=weight; new_b=weight/$\Delta$;
      if(bckt[$v$] > new_b) {
       bckt[$v$] = new_b; bckt_set = bckt_set $\cup$ {new_b};}
      if(new_b == b) {active[$v$]=true; bckt_empty=true;}}}}}
}
\end{lstlisting}

The algorithm works in epochs. In each epoch, a bucket $b$ is initialized with vertices 
whose tentative distances are $[(b-1)\Delta,b\Delta)$, and relaxations are computed until all vertices within distance $b\Delta$ are found.
This means that in epoch $b$, edges are relaxed only from vertices whose final distances are within $[(b-1)\Delta,b\Delta)$. 

Let $L$ be the maximum weighted distance between any pair of vertices in the graph,
and let $l_\Delta$ be the number of iterations done in any epoch.
If $n_i$ vertices fall into the $i$th bucket, at the $i$th epoch
$\mathcal O(l_\Delta\hat{d})$ executions of \krelax{$n_i$} will relax edges of vertices in the current bucket  
and up to $l_\Delta$ executions of \kfilter{$n_i$} will be used to update the set of vertices in the current bucket. 
So each edge will be relaxed $\mathcal O(l_\Delta)$ times.
There are a total of $L/\Delta$ epochs, so the complexity of $\Delta$-stepping is
(1) $\mathcal O((L/\Delta)l_\Delta(m/P+\hat{d}))$ time and $\mathcal O((L/\Delta)ml_\Delta )$ work using pulling,
(2) $\mathcal O(ml_\Delta/P+(L/\Delta)l_\Delta\hat{d})$ time and $\mathcal O(ml_\Delta)$ work using pushing in CRCW-CB, 
(3) $\mathcal O(\log(\hat{d}))$ more than (2) using pushing in CREW.
Pushing achieves a smaller cost, since we relax the edges leaving each node in only one of $L/\Delta$ epochs.
These results may be extrapolated to specific types of graphs considered in the original analysis~\cite{meyer2003delta}.


\macb{Conflicts}
In pushing, there is a write conflict for each of $O(ml_\Delta)$ edge relaxations.
In pulling, there is a read conflict for each of $O((L/\Delta)ml_\Delta)$ edge relaxations.

\macb{Atomics/Locks}
In pushing, each edge relaxation can be performed via a CAS atomic (in total $O(ml_\Delta)$ of these).

\macsubsection{Betweenness Centrality}
\label{sec:pram_bc}

BC is illustrated in Algorithm~\ref{lst:pram_bc}.
For each source vertex, we first compute a BFS to count the multiplicities of each shortest path and 
store all predecessors that are on some shortest path for each destination vertex.
The list of predecessors is then used to define a shortest path tree.
To calculate the partial centrality scores, this tree is traversed via BFS starting from the tree leaves.
We use the \texttt{ready} array to ensure tree-nodes enter the frontier only once the partial centrality
updates of all of their children are accumulated.

This algorithm (parallel Brandes) was described in detail~\cite{brandes2001faster, madduri2009faster}.
The approach is dominated by $2n$ BFS invocations, the cost of which is analyzed in~\cref{sec:pram_bfs}.
For directed graphs, SSSP (e.g., $\Delta$-stepping) must be used to compute each shortest-path tree.
Given the shortest-path tree the partial centrality scores can be computed via BFS
in the same way as for undirected graphs.
Computationally, the most significant difference of BC from SSSP and BFS, is the presence
of additional parallelism. Many source vertices can be processed independently, so up to
$O(n^2)$ processors can be used by running $n$ independent instances of BFS or SSSP.


\macb{Conflicts and Atomics/Locks}
The number of conflicts as well as atomics or locks matches that of BFS or SSSP
and can vary by the factor of up to $O(n)$ (depending on the amount of
additional parallelism). Yet, since the accumulation operator for the second
BFS uses floating point numbers, locks are required instead of atomics.  This
can be alleviated by maintaining sets of successors instead of predecessors as
proposed by Bader et al.~\cite{bader2007approximating}, which we identify as
another opportunity for using either pushing or pulling. We elaborate on it in
the technical report. 





\begin{lstlisting}[belowskip=-1.5em,aboveskip=0.5em,float=h!,label=lst:pram_bc,caption=(\cref{sec:pram_bc}) Push- and pull-based Betweenness Centrality.]
/* Input: a graph $G$. Output: centrality scores bc[1..$n$]. */

function BC($G$) { bc[1..$n$] = [$0$..$0$]
  Define $\Pi$ so that any $\Pi\ni u=(\text{index}_u, \text{pred}_u,\text{mult}_u,\text{part}_u)$;
  Define $u \acc_\text{pred}v$ with $u,v\in\Pi$ so that $u$ becomes $u=(\text{index}_u, \text{pred}_u\cup \text{index}_v, \text{mult}_u+\text{mult}_v, \text{part}_u)$;
  Define $u \acc_\text{part}v$ with $u,v\in\Pi$ so that $u$ becomes $u=(\text{index}_u, \text{pred}_u, \text{mult}_u, \text{part}_u+(\text{mult}_u/\text{mult}_v)(1+\text{part}_v))$;

  for $s \in V$ do [in par] {
    ready = $[1,\ldots, 1]$; ready[$s$] = 0;
    R = BFS($G$,ready,[$(1,\emptyset,0,0)..(s,\emptyset,1,0)..(n,\emptyset,0,0)$],$\acc_\text{pred}$)];
    Define graph $G'=(V,E')$ where $(u,v)\in E'$ iff $\text{index}_v\in\text{pred}_u$;
    Let ready[$u$] be the in-degree of $u\in V$ in $G'$;
    R = BFS($G'$,ready,R,$\acc_\text{part}$);
    for $(\text{index}_u, \text{pred}_u,\text{mult}_u,\text{part}_u) \in$ R do [in par] 
      bc[$u$] += $\text{part}_{u}$; }
\end{lstlisting}


\macsubsection{Boman Graph Coloring}
\label{sec:pram_bgc}


We present BGC in Algorithm~\ref{lst:pram_bgc}.
The algorithm proceeds for $L$ iterations, a quantity that is sensitive to both
the schedule of threads and the graph structure.
To limit the memory consumption, we bound the maximum count of
colors to $\mathcal{C}$. We use an opaque function \textsf{init}
that partitions $G$ and thus initializes the set of border vertices $\mathcal{B}$
and all the partitions $\mathscr{P} = \{\mathcal{P}_1 ... \mathcal{P}_s\}$.
The algorithm alternates between doing sequential graph coloring 
(\texttt{seq\_color\_partition}) and adjusting colors of bordering vertices.
The adjustment of colors of bordering vertices corresponds to an invocation of \krelax{$|\mathcal{B}|$}, in the worst case $|\mathcal{B}|=\Theta(n)$. 
Therefore, the complexity of BGC is
(1) $\mathcal O(L(m/P+\hat{d}))$ time and $\mathcal O(Lm)$ work using pulling,
(2) $\mathcal O(L(m/P+\hat{d}))$ time and $\mathcal O(Lm)$ work using pushing in CRCW-CB,
(3) $\mathcal O(\log(\hat{d}))$ more than (2) using pushing in CREW.


\macb{Conflicts}
Pushing/pulling require $\mathcal O(Lm)$ write/read conflicts.

\macb{Atomics/Locks}
In pushing and pulling the write conflicts can be resolved via CASes (a total of $\mathcal O(Lm)$ of these).

\begin{tikzpicture}[remember picture,overlay,pin distance=0cm]
\draw[fill=black!15, opacity=1, inner sep=4pt, rounded corners=4pt]
  ([shift={(-0.25em,0.55em)}]pic cs:bgc-ps) 
    rectangle
  ([shift={(25em,-0.6em)}]pic cs:bgc-ps);
\fill ([shift={(23.25em,0.0em)}]pic cs:bgc-ps) node[inner sep=2pt, rounded corners, text=white, fill=black!75, font=\tiny] {\macb{PUSHING}};
\draw[fill=black!15, opacity=1, inner sep=4pt, rounded corners=4pt]
  ([shift={(-0.25em,-0.6em)}]pic cs:bgc-ps) 
    rectangle
  ([shift={(25em,-1.75em)}]pic cs:bgc-ps);
\fill ([shift={(23.25em,-1.2em)}]pic cs:bgc-ps) node[inner sep=2pt, thin,rounded corners, text=white, fill=black!75, font=\tiny] {\textsf{\textbf{PULLING}}};
\end{tikzpicture}

\begin{lstlisting}[aboveskip=-1em, belowskip=-3.5em,float=h!, label=lst:pram_bgc, caption=(\cref{sec:pram_bgc}) Push- and pull-based Boman Graph Coloring.]
// Input: a graph $G$. Output: An array of vertex colors c[1..$n$].
// In the code, the details of functions seq_color_partition and
// init are omitted due to space constrains. 

function Boman-GC($G$) {
  done = false; c[1..$n$] = [$\emptyset$..$\emptyset$]; //No vertex is colored yet
  //avail[i][j]=1 means that color j can be used for vertex i.
  avail[1..$n$][1..$\mathcal{C}$] = [1..1][1..1]; init($\mathcal{B}, \mathscr{P}$);
  while (!done) {|\label{ln:core-cr-example-start}|
    for $\mathcal{P} \in \mathscr{P}$ do in par {seq_color_partition($\mathcal{P}$);}
    fix_conflicts(); } }|\label{ln:core-cr-example-end}|

function fix_conflicts() {
  for $v \in \mathcal{B}$ in par do {for $u \in N(v)$ do 
    if (c[$u$] == c[$v$]) {
|\tikzmark{bgc-ps}|      {avail[$u$][c[$v$]] = $\emptyset$ $\text{\encircle{W}}$ $\text{\enbox{i\bf}}$;}

|\tikzmark{bgc-pl}|      {avail[$v$][c[$v$]] = $\emptyset$ $\text{\encircle{R}}$ $\text{\enbox{i\bf}}$;}
  }}
\end{lstlisting}


\begin{tikzpicture}[remember picture,overlay,pin distance=0cm]
\draw[fill=black!10, opacity=1, inner sep=4pt, rounded corners=4pt]
  ([shift={(-0.25em,0.55em)}]pic cs:mst-ps) 
    rectangle
  ([shift={(25em,-3.4em)}]pic cs:mst-ps);
\fill ([shift={(23em,-0.2em)}]pic cs:mst-ps) node[rounded corners, text=white, fill=black!75, font=\tiny] {\macb{PUSHING}};
\draw[fill=black!10, opacity=1, inner sep=4pt, rounded corners=4pt]
  ([shift={(-0.25em,-3.4em)}]pic cs:mst-ps) 
    rectangle
  ([shift={(25em,-6em)}]pic cs:mst-ps);
\fill ([shift={(23em,-4.8em)}]pic cs:mst-ps) node[thin,rounded corners, text=white, fill=black!75, font=\tiny] {\macb{PULLING}};
\end{tikzpicture}

\begin{lstlisting}[belowskip=-2em,aboveskip=0em,float=h!,label=lst:pram_mst,caption=(\cref{sec:pram_mst}) Push- and pull-based Boruvka MST.]
function MST_Boruvka(G) {
 sv_flag=[1..$v$]; sv=[{1}..{$v$}]; MST=[$\emptyset$..$\emptyset$];
 avail_svs={1..$n$};	max_e_wgt=$\max_{v,w\in V}(\mathcal{W}_{(v,w)}+1)$;

 while avail_svs.size() > 0 do {avail_svs_new = $\emptyset$;
  for flag $\in$ avail_svs do in par {min_e_wgt[flag] = max_e_wgt;}
|\tikzmark{mst-part-1}|  for flag $\in$ avail_svs do in par {
   for $v \in$ sv[flag] do {
    for $w \in N(v)$ do [in par] {
|\tikzmark{mst-ps}|     if (sv_flag[$w$] $\neq$ flag) $\wedge$ 
        ($\mathcal{W}_{(v,w)}$ < min_e_wgt[sv_flag[$w$]]) $\text{\encircle{R}}$ {
      min_e_wgt[sv_flag[$w$]] = $\mathcal{W}_{(v,w)}$ $\text{\encircle{W}}$ $\text{\enbox{i\bf}}$;
      min_e_v[sv_flag[$w$]] = $w$; min_e_w[sv_flag[$w$]] = $v$ $\text{\encircle{W}}$ $\text{\enbox{i\bf}}$; 
      new_flag[sv_flag[$w$]] = flag  $\text{\encircle{W}}$ $\text{\enbox{i\bf}}$; }
|\tikzmark{mst-pl}|     if (sv_flag[$w$] $\neq$ flag) $\wedge$ ($\mathcal{W}_{(v,w)}$ < min_e_wgt[flag]) $\text{\encircle{R}}$ {
      min_e_wgt[flag] = $\mathcal{W}_{(v,w)}$; min_e_v[flag] = $v$;
      min_e_w[flag] = $w$; new_flag[flag] = sv_flag[$w$]; }$\text{\encircle{R}}$
 } } }
  while flag = merge_order.pop() do {
    neigh_flag = sv_flag[min_e_w[flag]];
    for $v$ $\in$ sv[flag] do sv_flag[flag] = sv_flag[neigh_flag];
    sv[neigh_flag] = sv[flag] $\cup$ sv[neigh_flag];
    MST[neigh_flag] = MST[flag] $\cup$ MST[neigh_flag] 
                      $\cup$ { (min_e_v[flag], min_e_w[flag]) }; } }
\end{lstlisting}

\macsubsection{Boruvka Minimum Spanning Tree}
\label{sec:pram_mst}

Push- and pull-based Boruvka is shown in Algorithm~\ref{lst:pram_mst}. Due to space constraints, it only displays pushing/pulling
when selecting the minimum edge adjacent to each supervertex.
The algorithm starts with $n$ supervertices and reduces their number by two at every iteration.
The supervertex connectivity graph can densify throughout the process with supervertices having degree $\Theta(n)$. 
However, the supervertices will always contain no more than $m$ edges overall.
Determining the minimum-weight edge for all supervertices requires $O(n^2/P)$ 
time and $O(m)$ work assuming each supervertex is processed sequentially.
Merging the vertices requires $O(\log(n))$ time and $O(n)$ work via a tree contraction~\cite{gazit1988optimal} (our implementation uses a more simplistic approach).
Merging the edges connected to each vertex can be done via $O(n)$ invocations of a \krelax{$k$}, where $k=O(n)$ at the first iteration
and then the bound decreases geometrically. 
Over all $\log(n)$ steps, the complexity of Boruvka is
(1) $\mathcal O(n^2/P)$ time and $\mathcal O(n^2)$ work using pulling,
(2) $\mathcal O(n^2/P)$ time and $\mathcal O(n^2)$ work using pushing in CRCW-CB,
(3) $\mathcal O(\log(n))$ more than (2) using pushing in CREW.

Theoretically, known PRAM algorithms for finding connectivity and minimal spanning forests~\cite{awerbuch1987new} are much faster in time complexity.
Still, our simple scheme is fairly efficient in practice as supervertex degree generally grows much slower than in the worst case.

\macb{Conflicts}
Pushing/pulling require $\mathcal O(n^2)$ write/read conflicts.

\macb{Atomics/Locks}
The write conflicts in pushing can be handled via CAS atomics (in total $\mathcal O(n^2)$ of them).

\macsubsection{Further Analytical Considerations}
\label{sec:more_th}

We discuss some further extensions to our cost analyses.
Please note that due to space constrains, several additional
analyses can be found in the technical report.

\macb{More Parallelism}
Our analysis considered parallelism with $P \leq \mathcal O(n)$.
However, our pseudocodes specify additional potential sources of parallelism in many of the algorithms.
Up to $m$ processors can be used in many cases (and even more for TC), but in this scenario, the distinction between pushing and pulling disappears.

\macb{Directed Graphs}
Pushing and pulling differ interestingly for directed graphs. 
Pushing entails iterating over all outgoing edges of a subset of the vertices,
while pulling entails iterating over all incoming edges of all (or most) of the vertices.
Thus, instead of $\hat{d}$ some cost bounds would depend on $\hat{d}_{out}$ and $\hat{d}_{in}$ for pushing and pulling, respectively;
more details are in the technical report.


\macsubsection{Discussion \& Insights}
\label{sec:insights_th}

We finally summarize the most important insights.

\macb{Write/Read Conflicts}
Pushing entails more write conflicts that must be resolved with locks or atomics (read conflicts
must be resolved only under the EREW model). An exception is BC where the difference
lies in the type of the data that causes conflicts (floats for pushing and integers
for pulling as was remarked in the past work~\cite{madduri2009faster}).
Moreover, traversals (BFS, BC (Part~2), SSSP) entail more read conflicts with pulling
(e.g., $\mathcal{O}(D n \hat{d})$ in the BFS based on pulling and none in the push-based BFS). 

\macb{Atomics/Locks}
We now summarize how conflicts translate into used atomics or
locks.
In many algorithms, pulling removes atomics or locks completely (TC, PR, BFS, $\Delta$-Stepping, MST).
In others (BC), it changes the type of conflicts from {\small$\text{\enbox{f\bf}}$} to
{\small$\text{\enbox{i\bf}}$}, enabling the utilization of atomics 
and removing the need for locks~\cite{intel64and}. 
%

\macb{Communication/Synchronization}
The above analyses show that pulling reduces 
synchronization compared to pushing (e.g., fewer atomics in TC). In contrast,
pushing limits communication (e.g., the number of memory reads in BFS).

\macb{Complexity}
Pulling in traversals (BFS, BC, SSSP-$\Delta$) entails more
time and work (e.g., see BFS).
On the other hand, in schemes such as PR that update all vertices
at every iteration, pulling avoid write conflicts. 
As a result, for PR and TC, pulling is faster than pushing in the PRAM CREW model by a logarithmic factor.

\macsection{Accelerating Pushing \& Pulling}
\label{sec:strategies}

Our analysis in~\cref{sec:th} shows that most push- and pull-based algorithms
entail excessive counts of atomics/locks and reads/writes, respectively.
We now describe strategies to reduce both. 
%


%


\macb{Partition-Awareness (PA, in Pushing)}
We first decrease the number of atomics by
\emph{transforming the graph representation to limit memory
conflicts}.
For this, we partition the adjacency array of each $v$ into two parts:
\emph{local} and \emph{remote}. The former contains the neighbors $u \in N(v)$
that are owned by $t[v]$ and the latter groups the ones owned by other threads.
All local and remote arrays form two contiguous arrays; offsets for each array
are stored separately.
This increases the representation size from $n+2m$ to $2n+2m$ but also enables
detecting if a given vertex $v$ is owned by the executing thread (to be updated
with a non-atomic) or if it is owned by a different thread (to be updated with
an atomic).
This strategy can be applied to PR, TC, and BGC.
Consider PR as an example. 
Each iteration has two phases. First, each thread updates its own vertices with
non-atomics. Second, threads use atomics to update vertices owned by other
threads.
Here, the exact number of atomics depends on the graph distribution and
structure, and is bounded by $0$ (if $\forall_{v \in V} \forall_{w \in N(v)}
t[v] \neq t[w]$) and $2m$ (if $\forall_{v \in V} \forall_{w \in N(v)} t[v] =
t[w]$). The former occurs if $G = (V,E)$ is bipartite (i.e., $V = U \cup W, U
\cap W = \emptyset$) and each thread only owns vertices from either $U$ or $W$.
The latter occurs if each thread owns all vertices in some $G$'s connected
component.
The number of non-atomics stays similar.
We show this example in Algorithm~\ref{lst:pa-example}.
The overhead from a barrier (line~\ref{ln:pa-bar}) is outweighed 
by fewer write conflicts (none in line~\ref{ln:pa-no-W}).

\begin{tikzpicture}[remember picture,overlay,pin distance=0cm]
\draw[fill=green!25, opacity=1, inner sep=4pt, rounded corners=4pt]
  ([shift={(0,0.7em)}]pic cs:pa-part-1) 
    rectangle
  ([shift={(25em,-2em)}]pic cs:pa-part-1);
\fill ([shift={(20.9em,0em)}]pic cs:pa-part-1) node[rounded corners, text=white, fill=black!75, font=\tiny] {\macb{PART 1: LOCAL UPDATES}};
\draw[fill=green!25, opacity=1, inner sep=4pt, rounded corners=4pt]
  ([shift={(0,0.7em)}]pic cs:pa-part-2) 
    rectangle
  ([shift={(25em,-1.9em)}]pic cs:pa-part-2);
\fill ([shift={(20.6em,0em)}]pic cs:pa-part-2) node[thin,rounded corners, text=white, fill=black!75, font=\tiny] {\macb{PART 2: REMOTE UPDATES}};
\end{tikzpicture}

\begin{lstlisting}[float=h,aboveskip=-1em, belowskip=-1em, label=lst:pa-example,caption=(\cref{sec:strategies}) Using Partition-Awareness for push-based PageRank.]
//The code below corresponds to lines |\ref{ln:pr-inner-start}|-|\ref{ln:pr-inner-end}| in Algorithm |\ref{lst:pram_pr}|.
//$V_L$ is a set of vertices owned by a local executing tread. 
//$V_G$ is a set of vertices owned by a tread different from the 
//local one. $V_L \cup V_G = V; V_L \cap V_G = \emptyset$.

|\tikzmark{pa-part-1}|for $v \in V_L$ do in par 
  for $u \in N(v)$ do [in par] 
    new_pr[$u$] += $(f\cdot$pr[$v$]$)/d(v)$ |\label{ln:pa-no-W}|
  
barrier(); //A lightweight barrier to synchronize all threads.|\label{ln:pa-bar}|

|\tikzmark{pa-part-2}|for $v \in V_G$ do in par 
  for $u \in N(v)$ do [in par] 
    new_pr[$u$] += $(f\cdot$pr[$v$]$)/d(v)$ $\text{\encircle{W}}$ $\text{\enbox{\textbf{i}}}$
\end{lstlisting}

\macb{Frontier-Exploit (FE, in Pushing/Pulling)}
The number of excessive reads/writes can be reduced by 
accessing only a fraction
of vertices in each iteration (the \emph{Frontier-Exploit} strategy), similarly to BFS. 
For example,
consider BGC.
In each iteration, every vertex is verified for potential
conflicts, entailing many memory reads, regardless of
whether pushing or pulling is used. 
To reduce the number of such reads,
a set of vertices $F \subseteq V$ that form a stable set (i.e., are not
neighbors) is selected at first and is marked with a specified color
$c_0$ (we denote different colors with $c_i, i \in \mathbb{N}$).
Then, the algorithm enters the main loop. In each iteration $i \ge 1$, all neighbors
of vertices in $F$ that have not yet been colored are assigned a color $c_i$; at the end of each iteration, 
$F$ is set to $\emptyset$ and
the newly marked neighbors
become the elements of $F$.
While iterating, for each vertex $v \in F$, if any of its neighbors $u \in
N(v)$ has the same color ($c_i$), then a conflict occurs and either $v$ or
$u$ (depending on the selected strategy) is assigned a color $c_{i+1}$ that was
not used before.  
This scheme resembles a BFS traversal with multiple sources selected at the
beginning and marked with a color $c_0$, and a frontier constituted
by vertices in $F$.
In pushing, the vertices in $F$ look for their uncolored
neighbors and mark them with $c_i$. In pulling, uncolored vertices look for
colored neighbors that are in $F$.

\begin{table*}
\footnotesize
\sf
\centering
\begin{tabular}{@{}l|lll|lll|ll|ll|ll|ll|ll|ll@{}}
\toprule
            & \multicolumn{3}{c}{\textbf{orc (PR)}} & \multicolumn{3}{c}{\textbf{rca (PR)}} & \multicolumn{2}{c}{\textbf{ljn (TC)}} & \multicolumn{2}{c}{\textbf{rca (TC)}} & \multicolumn{2}{c}{\textbf{orc (BGC)}} & \multicolumn{2}{c}{\textbf{rca (BGC)}} & \multicolumn{2}{c}{\textbf{pok (SSSP-$\Delta$)}} & \multicolumn{2}{c}{\textbf{rca (SSSP-$\Delta$)}}  \\
  Event     & Push & Push+PA & Pull & Push & Push+PA & Pull & Push & Pull & Push & Pull & Push & Pull & Push & Pull & Push & Pull & Push & Pull \\ \midrule
  L1 misses & 335M & 382M & 572M & 2,062M & 10,560M & 2,857M   & 10,815B & 10,684B  & 4,290M & 4,150M & 3,599B & 4,555B & 76,117M & 75,401M  & 54,57M  & 469M         & 11,01k & 76,19M \\
  L2 misses & 234M & 289M & 446M & 640k & 7,037M & 1,508M     & 700M & 645M & 2,303M & 2,215M & 3,656B & 4,418B & 74,48M & 73,92M            & 50,74M & 472M                           & 9,46k & 75,56M \\
  L3 misses & 64,75M  & 53,49M  & 181M & 348M & 537k & 866k       & 439M & 404M & 1,075M & 1,030M & 36,94M & 186M & 229k & 226k            & 8,52M & 11,43M                          & 308 & 279k \\
  TLB misses (data) & 130M & 142M & 129M & 12,21k & 274k & 21628 & 66,44M & 56,05M  & 37,45k   & 18,37k  & 229M & 411M & 4,046M & 3,801M &    3,763M       & 26,17M              & 403 & 513k \\
  TLB misses (inst) & 1188 & 336 & 1161 & 220 & 250 & 218                        & 1090 & 660 & 214     & 233 & 141k & 507k & 510 & 577 &     1,984k       &  11,22k              & 71 & 370 \\
  atomics   & 234M & 219M & 0 & 5,533M & 5,374M & 0 & 1,066B & 0 & 724k & 0 & 0 & 0 & 0 & 0 & 0 & 0 & 0 & 0 \\
  locks     & 0 & 0 & 0 & 0 & 0 & 0 & 0 & 0 & 0 & 0 & 219M & 219M & 5,358M & 5,358M & 902k & 44.60M  & 370 & 5.523M \\
  reads     & 1,196B & 1,183B & 1,187B & 43,39M & 62,59M & 37,49M & 3,169T & 3,158T & 158M & 135M & 17,90B & 23,04B & 419M & 404M & 2.435B & 2.339B & 42,32k & 454M \\
  writes    & 474M & 460M & 237M & 14.99M & 14,86M & 7,499M & 10,71B & 1,066B & 18,97M  & 725k & 3,866B & 4,201B & 97,44M & 96,95M & 718M & 663M & 9,545k & 100M \\
  branches (uncond) & 234M & 222M & 1971 & 5,533M & 7,340M  & 533 & 8,585B & 616k & 19,48M  & 631 & 2,714B & 2,902B & 67,58M & 67,40M & 441M & 421M & 5,171k & 64.3M \\
  branches (cond) & 474M & 466M & 240M & 15M & 18,79M & 9.467M & 3,173T & 3,173T & 156M & 156M & 23,62B & 32,46B & 524M & 495M & 2.27B & 2,192B & 35,03k & 518M \\ \bottomrule
                            \end{tabular}
\caption{(\cref{sec:sm_perf}) PAPI events for PR, BGC (average per iteration), and TC, SSSP-$\Delta$ (total count) for the SM setting (Daint, XC30, $T=16$).}
\label{tab:papi}
\end{table*}

\macb{Generic-Switch (GS, in Pushing/Pulling)}
%
%
Next, we use the idea of switching between pushing and pulling;
we want to 
not only reduce communication, but also
\emph{limit the iteration count}. We refer to the strategy as
\emph{Generic-Switch}.
As an example, consider the above-described BGC enhanced with
Frontier-Exploit.
%
%
Pushing itself results in the excessive number of
iterations.
This is because, when the number of vertices to be colored is low (our
experiments indicate $<0.1n$), threads often conflict with each other,
requiring more iterations.
Switching to pulling may prevent new iterations as no conflicts are generated.
Yet, using pulling too early would entail excessive memory accesses
(few vertices are colored).
Thus, 
one must carefully select a switching moment or strategy, for
example switch if the ratio of the number of the colored
vertices to the generated conflicts (in a given iteration) exceeds a certain
threshold.




\macb{Greedy-Switch (GrS, in Pushing/Pulling)}
Generic-Switch not always brings the desired speedups.
For example, BGC with Frontier-Exploit may still need many iterations to color a
small fraction of the remaining vertices due to many conflicts 
between threads that share vertices.
In such cases, it is more advantageous to completely switch from a parallel
variant (regardless of whether it does pushing or pulling) to an optimized
greedy scheme.


\macb{Conflict-Removal (CR, Pushing/Pulling)}
The final strategy (see Algorithm~\ref{lst:ci-example}) completely removes conflicts in both pushing and pulling.
Consider BGC as an example. Instead of solving conflicts over border
vertices (the set $\mathcal{B}$) in each iteration, one can first use an
optimized scheme (e.g., greedy sequential) to color them without any conflicts
(thus, this scheme is advantageous if $|\mathcal{B}|$ is small compared
to $|V|$).
The remaining vertices can then be colored in parallel; no conflicts occur
either as every $v \in \mathcal{B}$ is already colored. 

\begin{lstlisting}[aboveskip=0em,float=h!, label=lst:ci-example, caption=(\cref{sec:strategies}) Example of Conflict-Removal with BGC.]
//The code below corresponds to lines |\ref{ln:core-cr-example-start}|-|\ref{ln:core-cr-example-end}| in Algorithm |\ref{lst:pram_bgc}|.
seq_color_partition($\mathcal{B}$)
for $\mathcal{P} \in \mathscr{P}$ do in par {seq_color_partition($\mathcal{P}$);} 
\end{lstlisting}

\macsection{Performance Analysis}
\label{sec:sh_perf_anal}

Finally, we investigate the performance of push/pull variants 
and the described acceleration strategies.
Due to a large amount of data we present and discuss in detail a small
representative subset; the remainder is in the report (see the link on
page~1).


\macb{Selected Benchmarks \& Parameters}
We consider the push- and pull-based variants,
strategies from~\cref{sec:strategies},
strong- and weak-scaling, Hyper-Threading (HT),
and static/dynamic OpenMP scheduling.
Two types of synthetic graphs are used: power-law
Kronecker~\cite{leskovec2010kronecker} and
Erdős-Rényi~\cite{erdHos1976evolution} graphs with $n \in \{2^{20}, ...,
2^{28}\}$ and $\overline{d} \in \{2^1, ..., 2^{10}\}$. We also use real-world
graphs (Table~\ref{tab:graphs}) of various
sparsities: low $\overline{d}$ and large $D$ (road networks), low
$\overline{d}$ and $D$ (purchase graphs), and large $\overline{d}$ with 
low $D$ (communities).
The graphs have up to 268M vertices and 4.28B
edges.

\begin{table}[!h]
\centering
\footnotesize
\sf
\begin{tabular}{@{}l|lrrrr@{}}
\toprule
\textbf{Type}                                                                           & \multicolumn{1}{l}{\textbf{ID}} & \multicolumn{1}{r}{\textbf{$n$}} & \multicolumn{1}{r}{\textbf{$m$}} & \multicolumn{1}{r}{\textbf{$\bar{d}$}} & \multicolumn{1}{r}{\textbf{$\bar{D}$}} \\ \midrule
\multirow{1}{*}{R-MAT graphs}          & rmat                                    & 33M-268M                          & 66M-4.28B          &   2-16  & 19-33           \\ \midrule
\multirow{2}{*}{Social networks}     
                                                                                        & orc                                 & 3.07M                          & 117M          &      39      & 9    \\ 
                                                                                        & pok                                 & 1.63M                          & 22.3M          &      18.75   &  11     \\\midrule
\multirow{1}{*}{Ground-truth~\cite{yang2015defining} community}     
                                                                                        & ljn                                 & 3.99M                          & 34.6M          &    8.67   & 17       \\ \midrule
\multirow{1}{*}{Purchase network} 
                                                                                        & am                                   & 262k                           & 900k           &   3.43  & 32    \\ \midrule
\multirow{1}{*}{Road network}          & rca                                   & 1.96M                          & 2.76M          &   1.4  & 849           \\ \bottomrule
\end{tabular}
\caption{(\cref{sec:sh_perf_anal}) The analyzed graphs with skewed degree distributions.}
\label{tab:graphs}
\end{table}

\macb{Used Programming Models}
We use threading to harness SM systems. For DM 
machines, we use Message
Passing (MP, also denoted as Msg-Passing) and Remote Memory Access (RMA)~\cite{gerstenberger2013enabling}.
In MP, processes communicate explicitly and synchronize implicitly with
messages~\cite{besta2015accelerating}. In RMA, processes communicate and synchronize explicitly by accessing
remote memories with puts, gets, or atomics, and ensuring consistency with
flushes~\cite{gerstenberger2013enabling, besta2015active, besta2014fault}.

\macb{Counted Events}
We incorporate the total of nine performance counters for detailed analyses
of: cache misses (L1, L2, L3), reads and writes,
conditional/unconditional branches, and data/instruction TLB misses. We also
manually count issued atomics~\cite{schweizer2015evaluating} and acquired locks~\cite{schweizer2015evaluating}. 
Memory operations and cache/TLB misses are important as many graph algorithms
are memory-bound~\cite{beamer2015gail}. Branches were also shown to impact
performance in graph processing~\cite{green2014branch}.
Finally, in distributed settings we count sent/received messages, issued
collective operations, and remote reads/writes/atomics.

\macb{Experimental Setup and Architectures}
We use the following systems to cover various types of machines:

%
\begin{itemize}[leftmargin=0.5em,noitemsep]
%
%
\item \macb{Cray XC nodes} from the CSCS supercomputing systems. 
We use XC50 and XC40 nodes from the Piz Daint machine. An XC50
node contains a 12-core Intel Xeon E5-2690 CPU with 64 GiB
RAM. Each XC40 node contains an 18-core Intel Xeon E5-2695 CPU with
64 GiB RAM. We also show results for XC30 nodes (with an 8-core Intel
E5-2670 Sandy Bridge CPU and 32 GiB RAM) 
from a past Daint version.
Finally, we also provide results for XC40 nodes from a past Piz Dora
system (referred to as XC40*); they contained 12-core Intel Haswells E5-2690 and 64 GiB RAM.
All nodes are HT-enabled. The interconnection~\cite{besta2014slim} in all the cases is based on
Cray's Aries and it implements the Dragonfly topology~\cite{dally08}. 
%
%
This machines represents massively parallel HPC systems.
%
%
%
%
\item \macb{Trivium V70.05} is a server with Intel Core i7-4770 
(with four 3.4 GHz Haswell 2-way multi-threaded cores). Each core has 32 KB of L1
and 256 KB of L2 cache. The CPU has 8 MB of shared L3 cache and 8 GB of RAM.
This option represents commodity machines.
\end{itemize}

\macb{Infrastructure and Implementation Details}
We use the PAPI library (v5.4.1.1) to access performance counters.
We spawn one MPI
process per core (or per one HT resource if applicable). We use Cray-mpich
(v.7.2.2) for MP and the foMPI library (v0.2.1)~\cite{gerstenberger2013enabling}
for RMA.
We also use OpenMP 4.0 and TBB from the Intel Programming Environment 6.0.3.
We compile the code (with the -O3 flag) with g++ v4.9.2 (on Trivium) and
Cray GNU 5.2.40 g++ (on CSCS systems).
The information refers to the current Daint system; others are covered 
in the technical report.


\macsubsection{Shared-Memory Analysis}
\label{sec:sm_perf}

We first analyze the differences in the SM setting.
The representative PAPI data for selected schemes is in Table~\ref{tab:papi}.
For each scheme, we discuss in more detail the results for graphs
with: large $\overline{d}$ and low $D$, and low $\overline{d}$ and
large $D$.

\macb{PageRank}
PR results can be found in Table~\ref{tab:pr_analysis}. 
In graphs with both high $\overline{d}$ (orc, ljn, poc) and low
$\overline{d}$ (rca, am), pulling outperforms pushing by $\approx$3\% and
$\approx$19\%, respectively.
The former requires no atomics, but its speedup is moderate as it also
generates more cache misses and branches as it accesses various neighbors,
requiring more random memory reads.  

\setlength{\tabcolsep}{2pt}

\begin{table}[!h]
\footnotesize
\sf
\centering
\begin{tabular}{@{}l|lllll|lllll@{}}
\toprule
                                & \multicolumn{5}{c}{\textbf{PageRank} [ms]}               & \multicolumn{5}{c}{\textbf{Triangle Counting} [s]} \\
                            $G$ & orc & pok & ljn & am & rca & orc & pok & ljn & am & rca \\ 
                            \midrule
                            Pushing & 572 & 129 & 264 & 4.62 & 6.68 & 11.78k & 139.9 & 803.5 & 0.092 & 0.014 \\
                            Pulling & 557 & 103 & 240 & 2.46 & 5.42 & 11.37k & 135.3 & 769.9 & 0.083 & 0.014 \\ 
                            \bottomrule
                            \end{tabular}
\caption{(\cref{sec:sm_perf}) Time per iteration for PageRank [ms] and the total time to compute for Triangle Counting [s] (SM setting, Daint, XC30, $T=16$).}
\label{tab:pr_analysis}
                            \end{table}

\setlength{\tabcolsep}{1pt}

\macb{Triangle Counting}
We now proceed to TC (Table~\ref{tab:pr_analysis}).
Large amounts of time are due to the high computational complexity
(\cref{sec:pram_tc});
this is especially visible in graphs with high $\overline{d}$.
Here, pulling always outperforms pushing (by $\approx$4\% for orc and
$\approx$2\% for rca). This is due to atomics but also
more cache misses caused by atomics.

\macb{Graph Coloring}
The BGC results are presented in Figure~\ref{fig:bgc_analysis}.
Pushing is always faster than pulling (by $\approx$10\% for orc and
$\approx$9\% for rca for iteration~1). More detailed measurements indicate that
the number of locks acquired is the same in both variants, but pushing always
entails fewer cache/TLB misses and issued reads and writes.

\begin{figure}[!h]
\centering
 \subfloat[The orc graph.]{
  \includegraphics[width=0.14\textwidth]{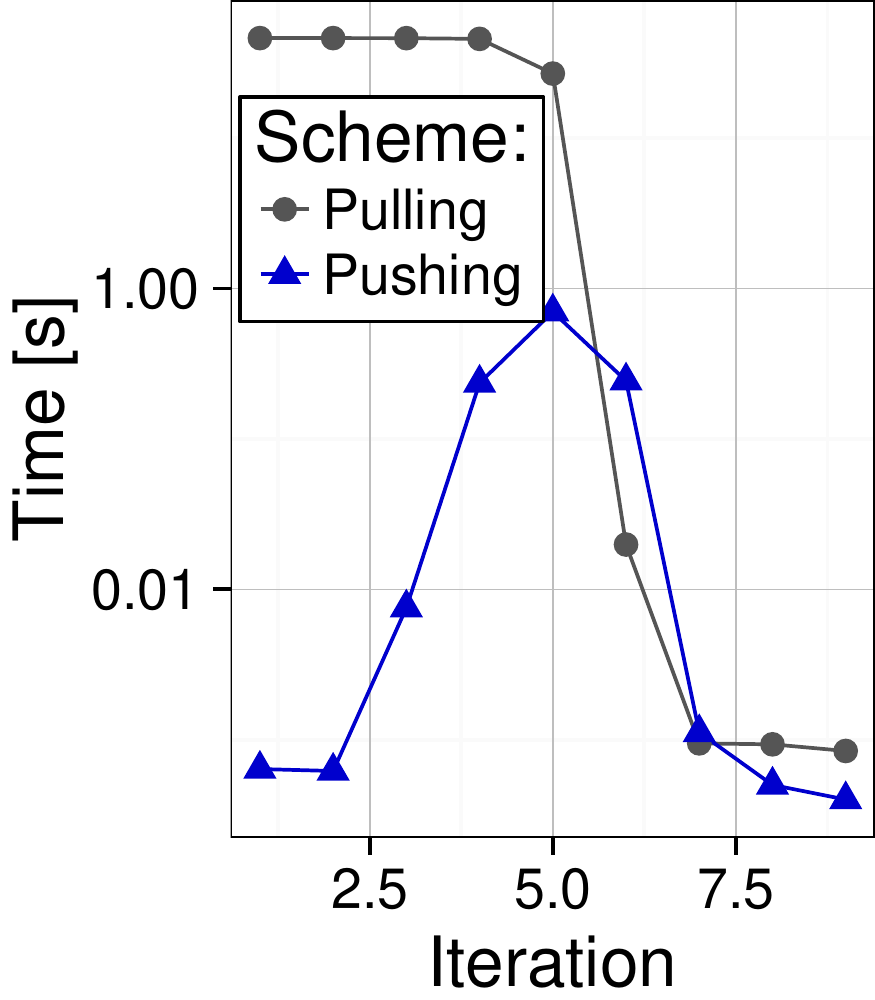}
  \label{fig:sssp-d_orc}
 }
 \subfloat[The am graph.]{
  \includegraphics[width=0.14\textwidth]{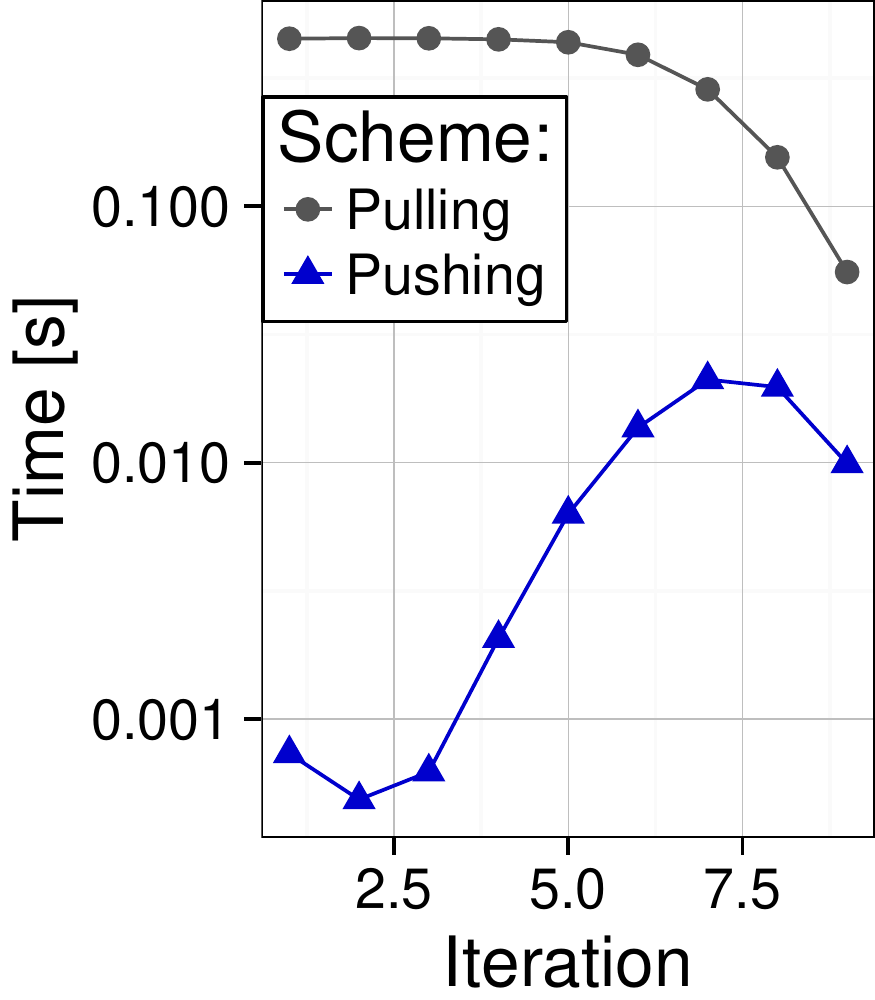}
  \label{fig:sssp-d_am}
 } 
 \subfloat[Varying ${\Delta}$ (orc).]{
  \includegraphics[width=0.14\textwidth]{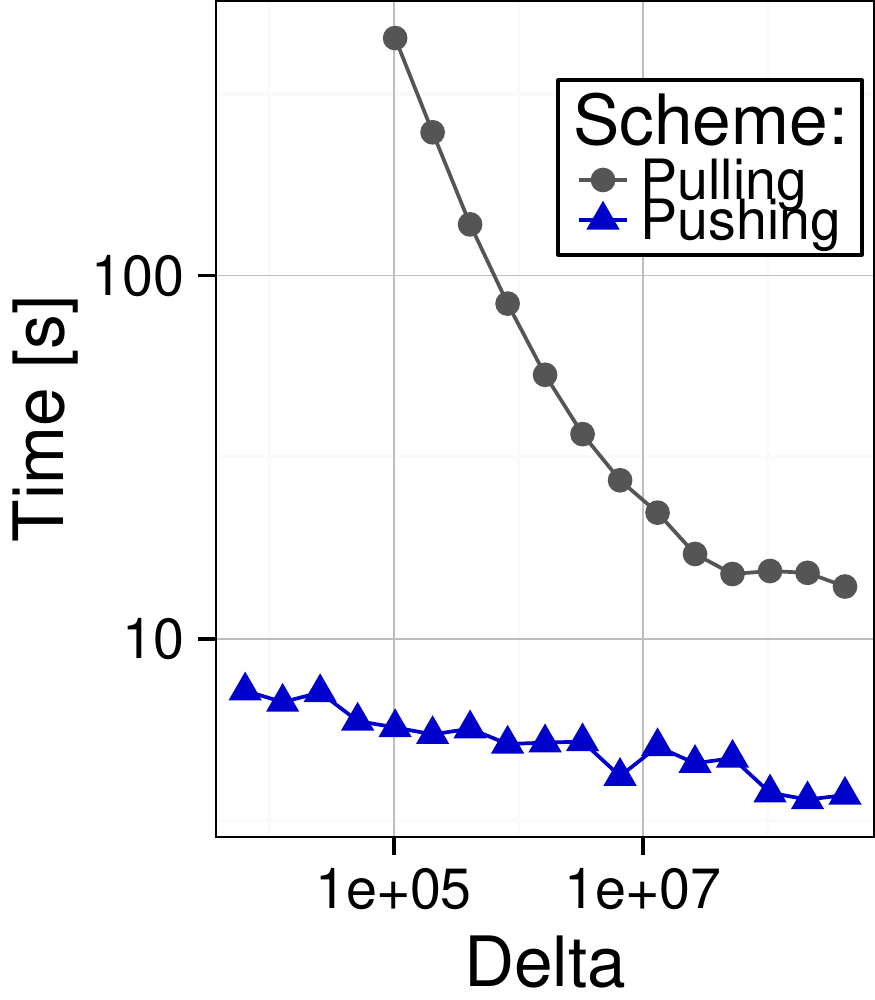}
  \label{fig:vary-delta}
 } 
\caption{(\cref{sec:sm_perf}) SSSP-$\Delta$ SM analysis (XC30, $T=16$).}
\label{fig:sssp-d_analysis}
\end{figure}

\macb{$\Delta$-Stepping}
The outcomes for orc and am can be found in Figure~\ref{fig:sssp-d_analysis}). Both push and pull
variants use locks. Yet, a higher number of memory accesses issued in most
iterations in the pull-based scheme limits performance. As expected, the
difference decreases after several iterations because the frontier grows (with pushing),
requiring more memory accesses. This is
especially visible in graphs with high $\overline{d}$ where pulling outperforms
pushing (e.g., iteration~6 for orc).
Moreover, illustrate in Figure~\ref{fig:vary-delta} that the larger $\Delta$ is,
the smaller the difference between pushing and pulling becomes.

\begin{figure*}[!t]
\centering
 \subfloat[PR, orc.]{
  \includegraphics[width=0.15\textwidth]{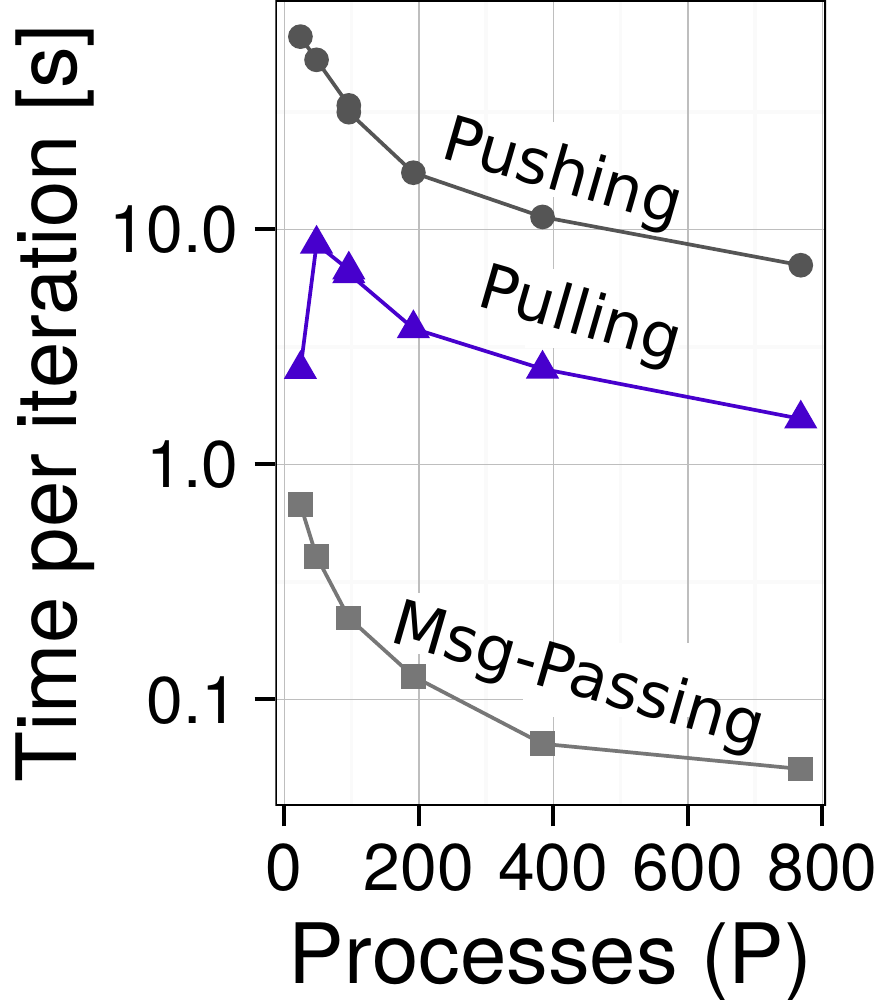}
  \label{fig:pr-dm_orc}
 }
 \subfloat[PR, ljn.]{
  \includegraphics[width=0.15\textwidth]{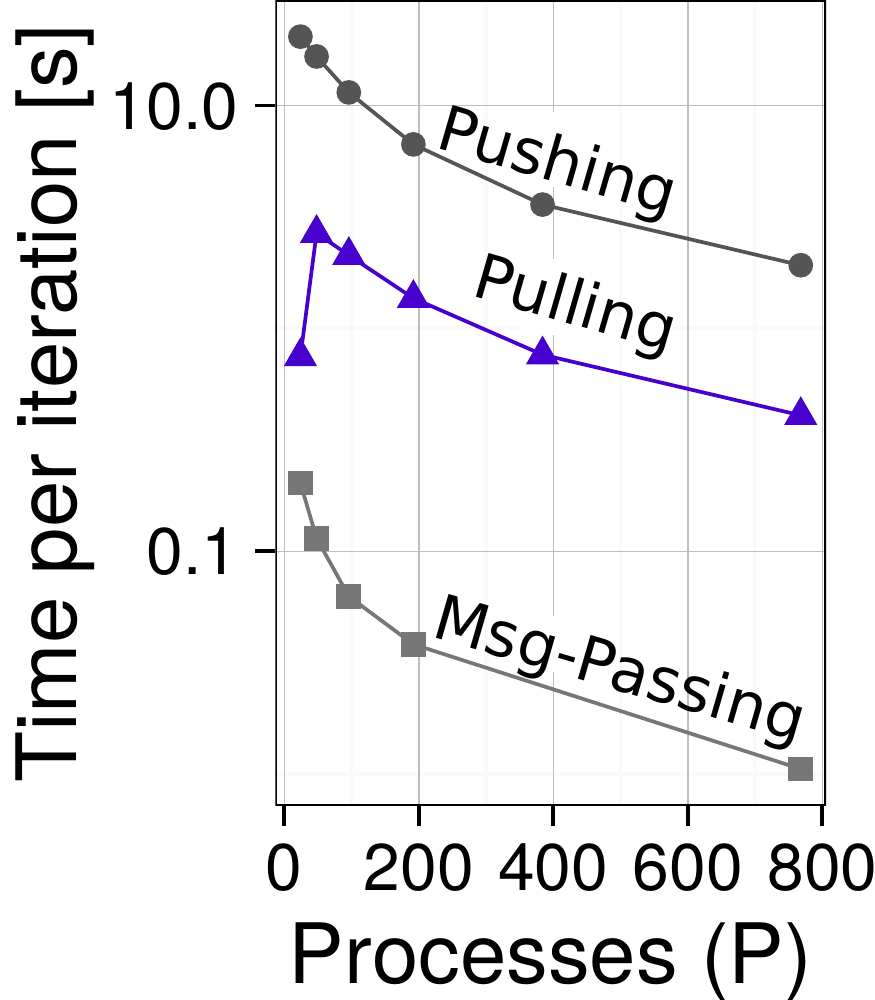}
  \label{fig:pr-dm_pok}
 } 
 \subfloat[PR, rmat, $n = 2^{25}$.]{
  \includegraphics[width=0.15\textwidth]{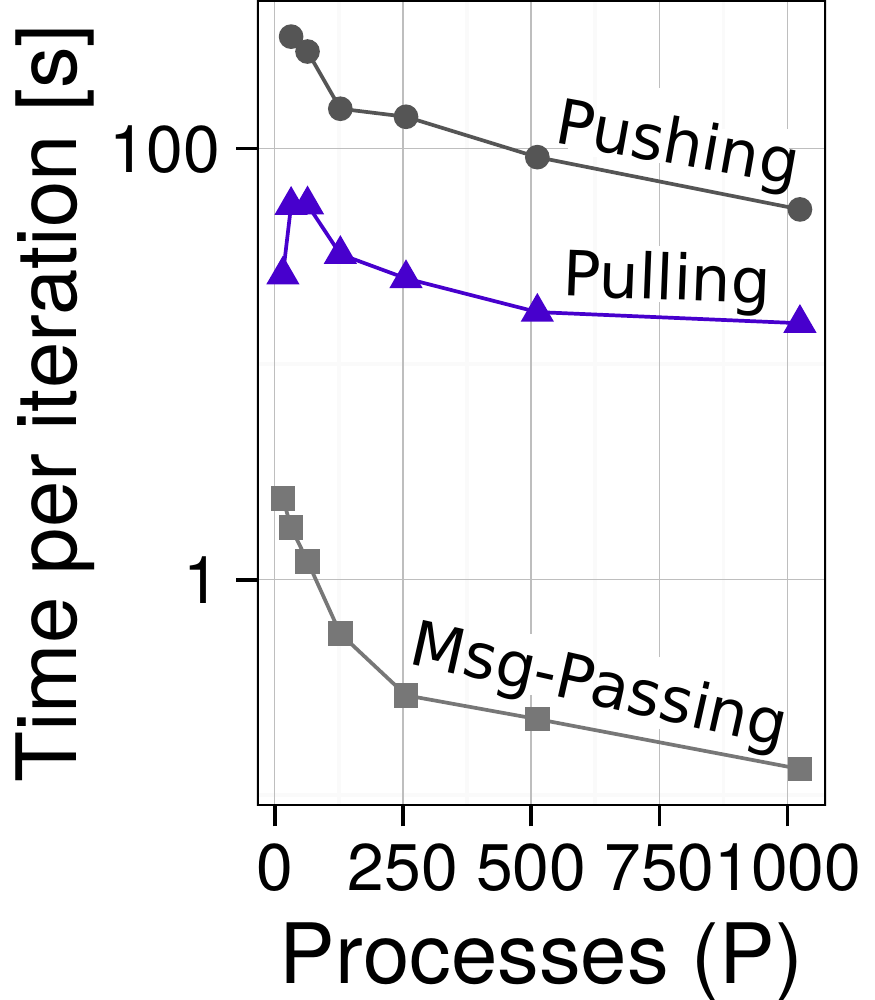}
  \label{fig:rmat_25}
 }
 \subfloat[PR, rmat, $n = 2^{27}$.]{
  \includegraphics[width=0.15\textwidth]{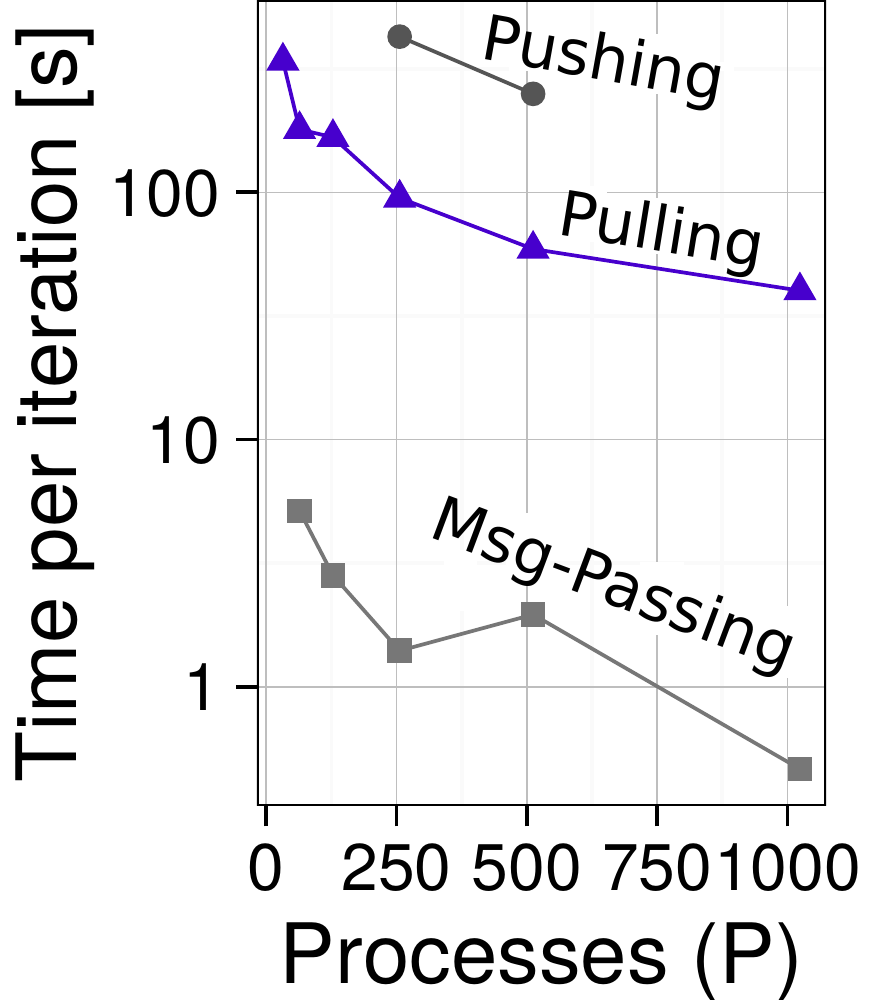}
  \label{fig:rmat_27}
 }
 \subfloat[TC, orc.]{
  \includegraphics[width=0.15\textwidth]{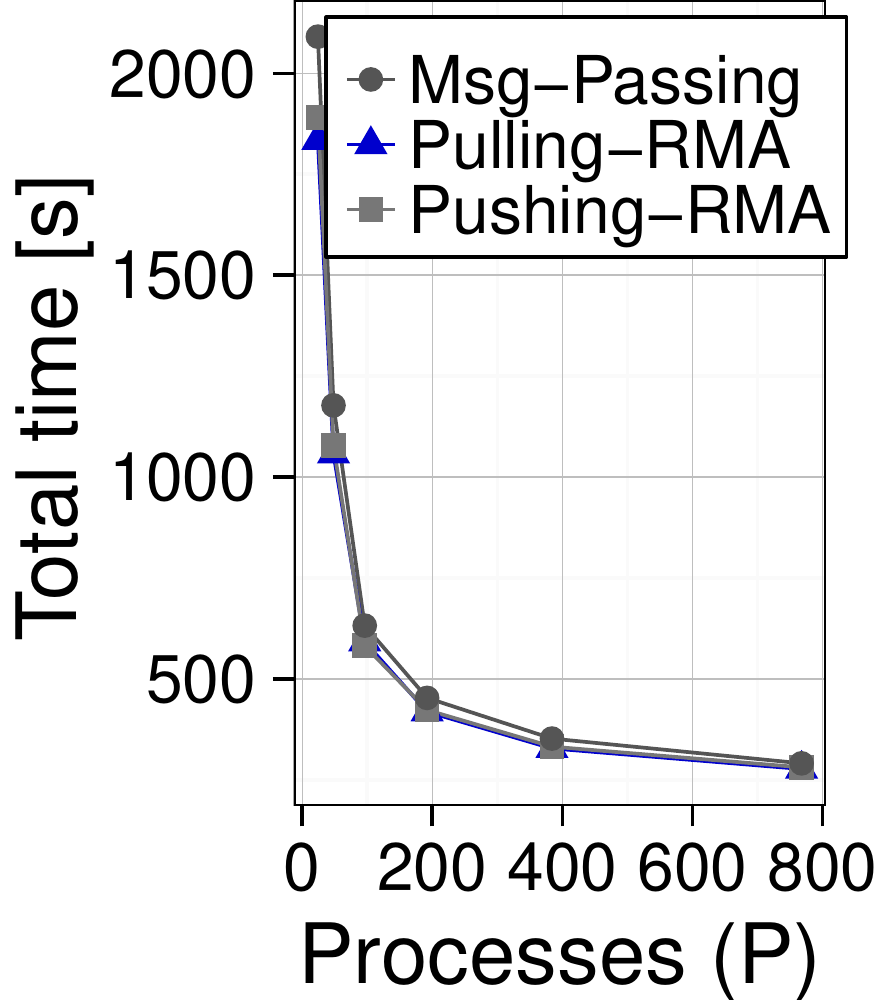}
  \label{fig:tc-dm_orc}
 } 
 \subfloat[TC, ljn.]{
  \includegraphics[width=0.15\textwidth]{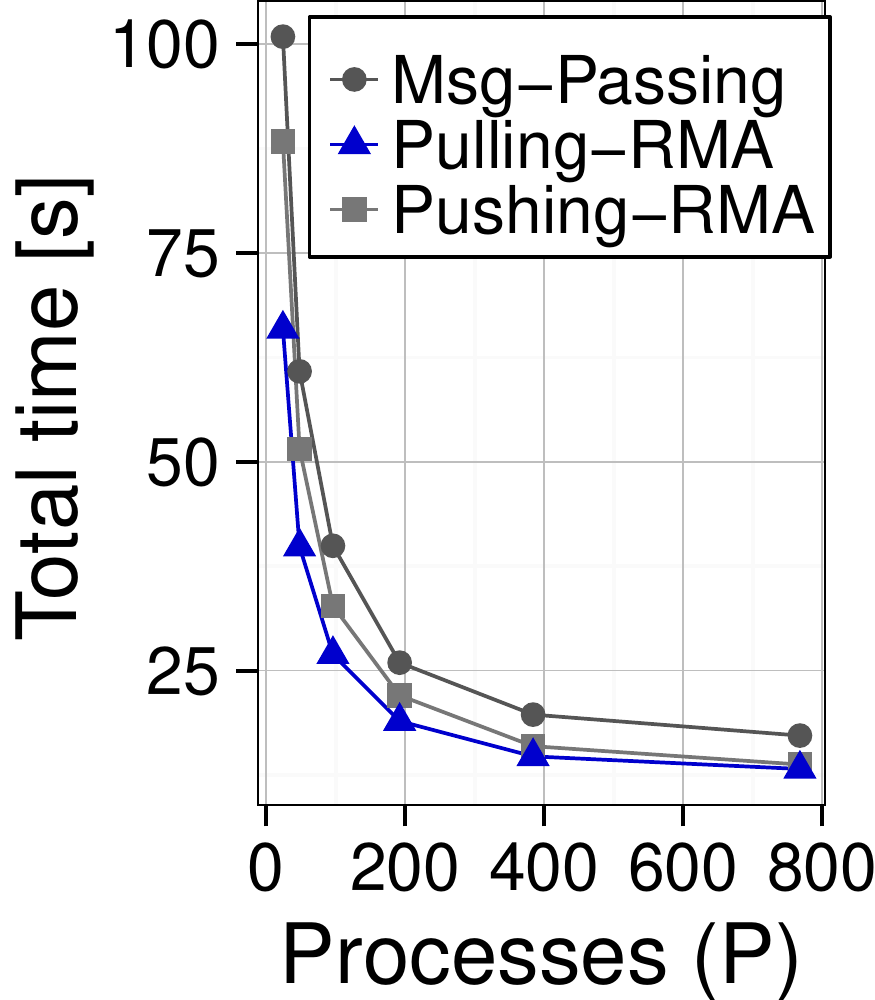}
  \label{fig:tc-dm_ljn}
 } 
\caption{(\cref{sec:dm-perf}) The results of the scalability analysis in the DM setting, strong scaling (rmat graphs: XC40, $T=24$; real-world graphs: XC40*, $T=24$).}
\label{fig:dm_analysis}
\end{figure*}

\macb{Breadth-First Search}
The results are similar to SSSP-$\Delta$;
pushing outperforms pulling in most cases. This is most
visible for rca (high $D$, low $\overline{d}$) due to many 
memory accesses.

\begin{figure}[!h]
\centering
 \subfloat[``Find Minimum''.]{
  \includegraphics[width=0.15\textwidth]{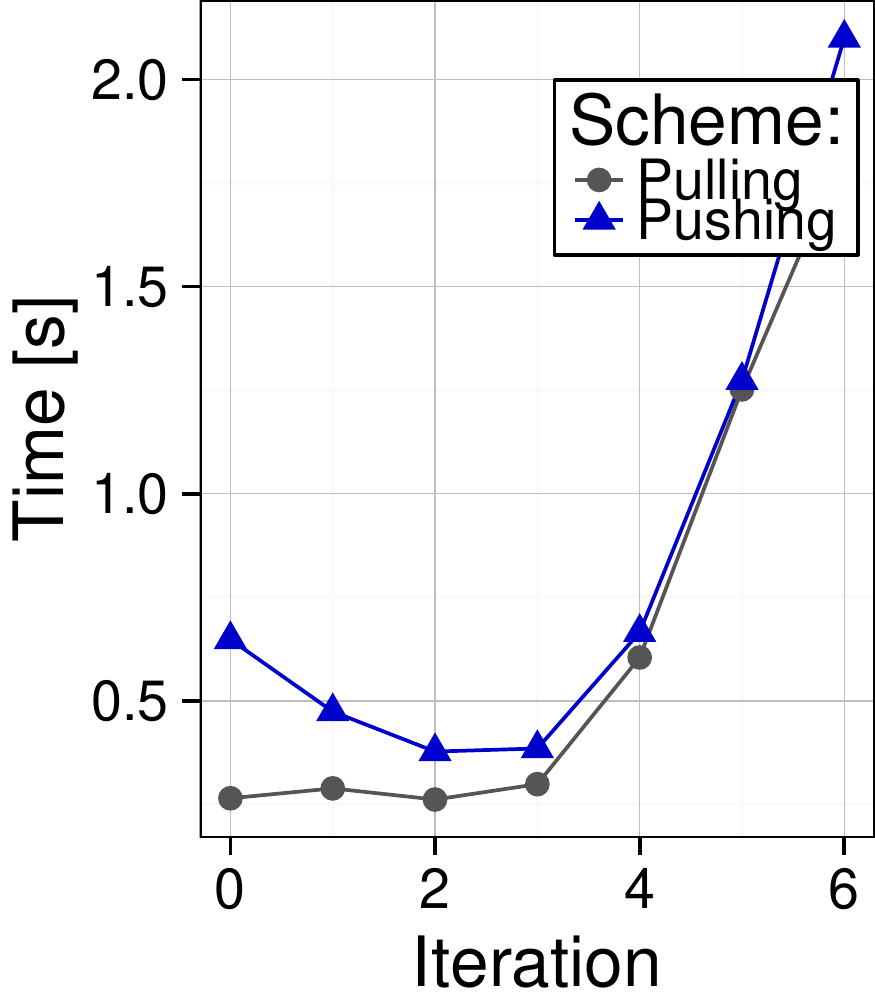}
  \label{fig:mst-orc}
 }
 \subfloat[``Build Merge Tree''.]{
  \includegraphics[width=0.15\textwidth]{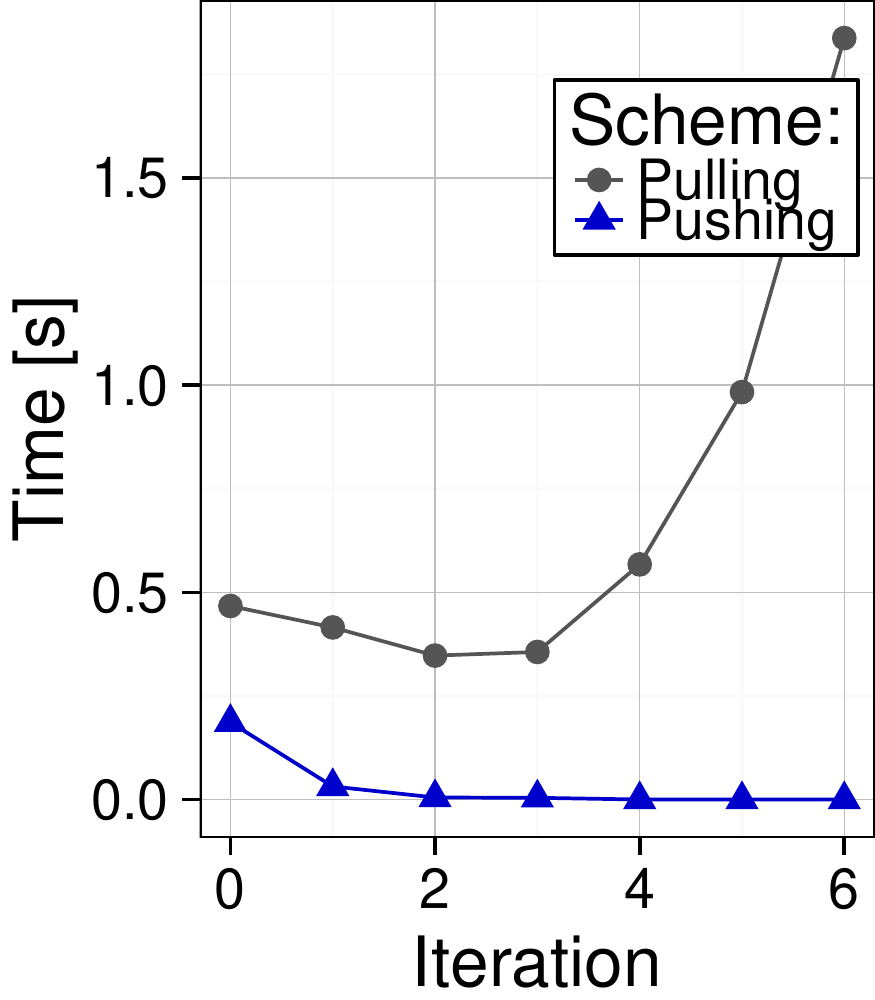}
  \label{fig:mst-orc}
 }
 \subfloat[``Merge''.]{
  \includegraphics[width=0.15\textwidth]{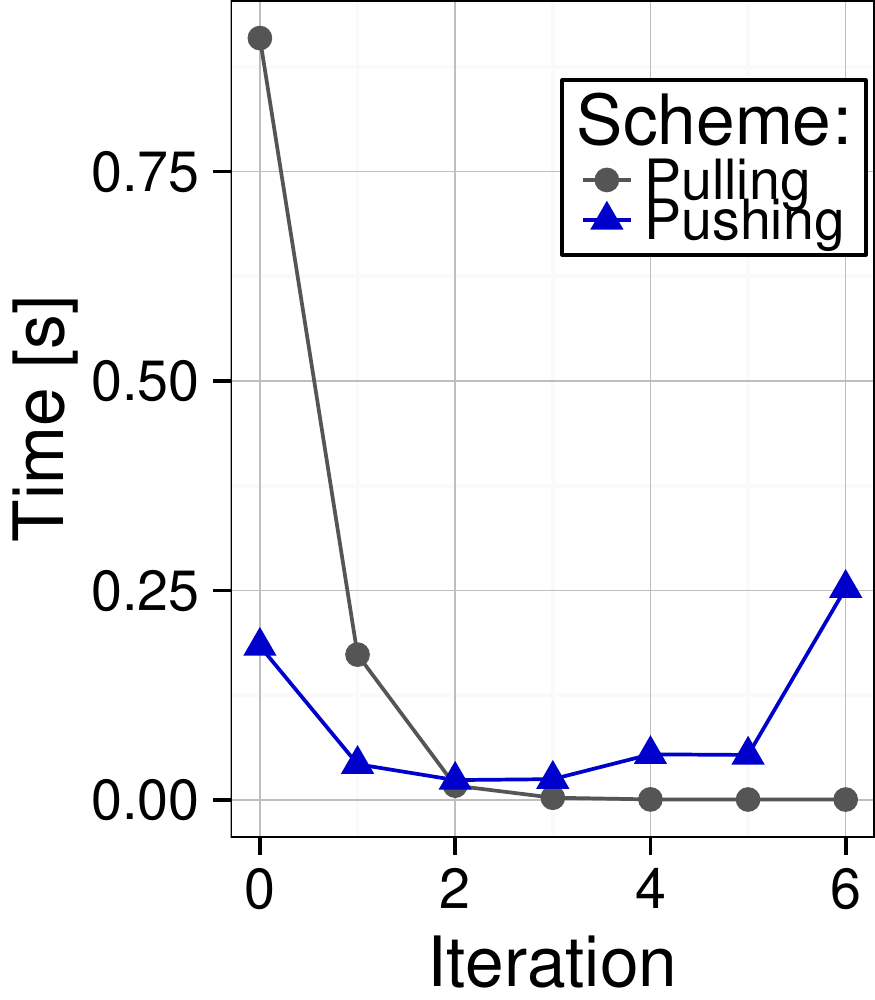}
  \label{fig:mst-orc}
 }
 %
\caption{(\cref{sec:sm_perf}) Illustration of the MST analysis, each subplot relates to a different phase (XC40, HT enabled, the orc graph, $T=16$).}
\label{fig:mst-res}
\end{figure}

\macb{Minimum Spanning Trees}
We illustrate the MST results in Figure~\ref{fig:mst-res}.
We analyze time to complete each of the three most time-consuming phases
of each iteration: Find Minimum (FM; looking for minimum-weight edges),
Build Merge Tree (BMT; preparing metadata for merging), and Merge
(M; merging of subtrees).
Now, pushing is faster than pulling in BMT and comparable in M. Yet,
it is slower in the most computationally expensive FM.
In summary, performance trends are similar to those of TC:
pushing is consistently slower ($\approx$20 for $T=4$) than pulling. 
This is because the latter entails no expensive write conflicts.

\begin{figure}[!h]
\centering

  \subfloat[First BFS runtime]{
   \includegraphics[width=0.15\textwidth]{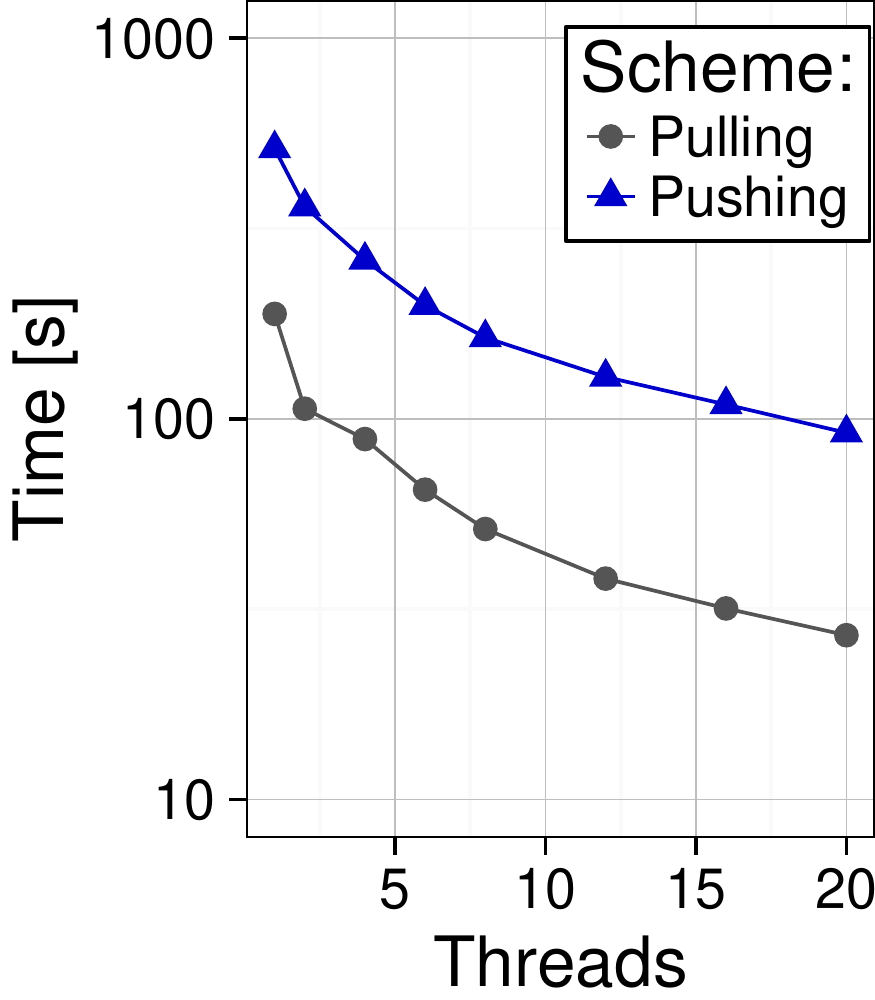}
   \label{fig:mst-orc}
  }
  \subfloat[Second BFS runtime]{
   \includegraphics[width=0.15\textwidth]{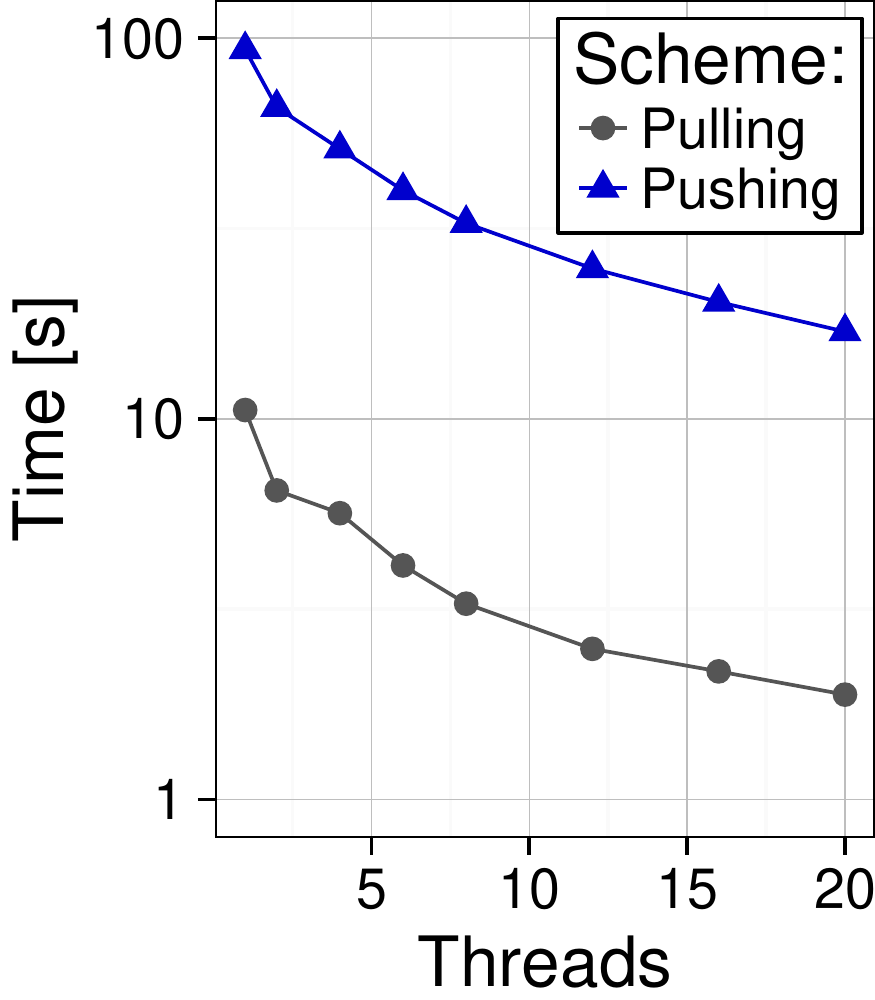}
   \label{fig:mst-orc}
  }
  \subfloat[Total runtime.]{
   \includegraphics[width=0.15\textwidth]{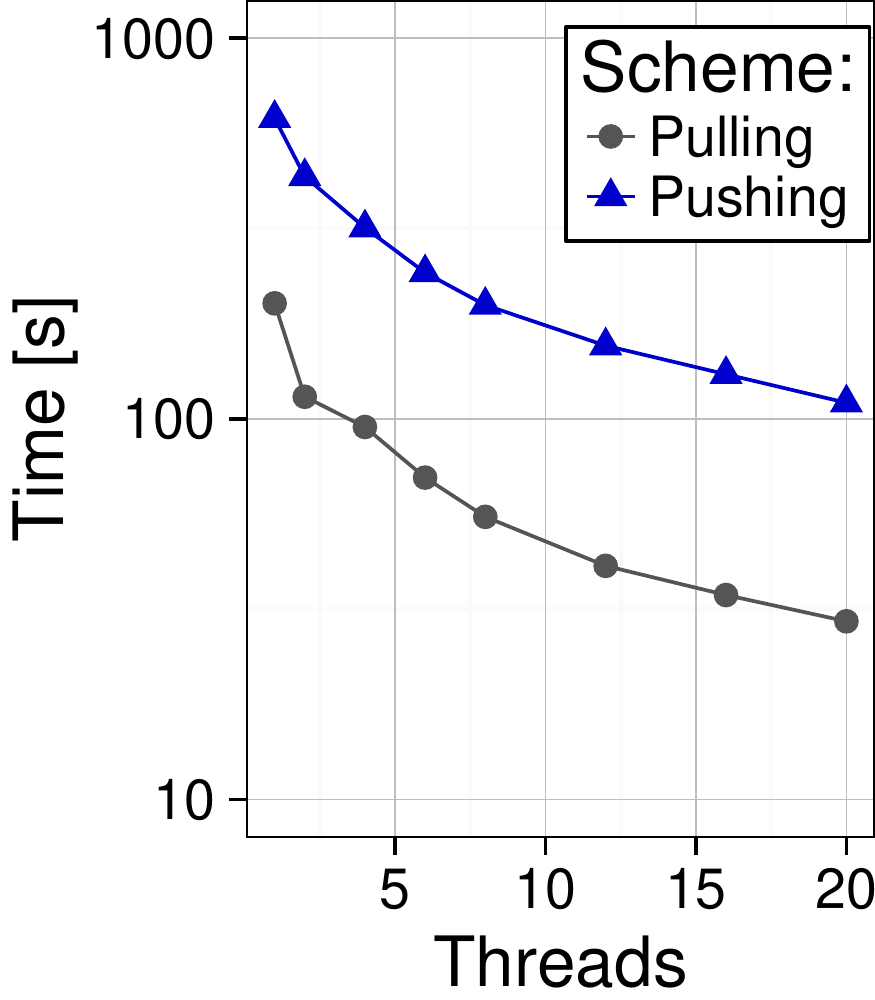}
   \label{fig:mst-orc}
  }
 %
\caption{(\cref{sec:sm_perf}) Illustration of the BC analysis (scalability, XC40, HT enabled, the orc graph, $T=16$).}
\label{fig:bc-res}
\end{figure}

\macb{Betweenness Centrality}
The results for BC can be found in Figure~\ref{fig:bc-res}.
We present the running times of both BFS traversals and the total BC runtime.
In each case, pushing is slower than pulling because of the higher amount
of expensive write conflicts that entail more synchronization in both
BC parts.

\macsubsection{Acceleration Strategies}
\label{sec:acc_perf}


We now evaluate the acceleration strategies (\cref{sec:strategies}).

\macb{Partition-Awareness (PA)}
We start with adding PA to PR (Table~\ref{tab:acc_pr_perf}).
In graphs with higher $\overline{d}$ (orc, ljn, poc), pushing+PA outperforms
pulling (by $\approx$24\%). This is because PA decreases atomics
(by 7\%) and comes with fewer cache misses ($\approx$30\%
for L1, $\approx$34\% for L2, and $\approx$69\% for L3) than pulling.
In sparser graphs (rca, am), surprisingly pushing+PA is the slowest
($\approx$205\% than pushing). This is because fewer atomics issued in
pushing+PA ($\approx$4\%) are still dominated by more branches ($\approx$23\%),
reads ($\approx$44\%), and cache misses ($\approx$53\% for L3).
We conjecture that in graphs with high $\overline{d}$, PA 
enhances pushing as the latter entails more atomics that dominate the performance.
This is visible as both variants reduce the number of cache misses if adjacency lists
are long and use better cache prefetchers. 
Then, for low $\overline{d}$, adjacency lists are short on average, 
giving more cache misses in pushing+PA and pushing, making pulling the
fastest. The worst performance of pushing+PA is due to the synchronization
overheads (it splits each iteration into two phases separated by a barrier)
that are no longer compensated with more effective cache utilization.

\macb{Frontier-Exploit (FE), Generic/Greedy-Switch (GS/GrS)}
We now apply these strategies to BGC, ensuring the same number of colors for
each coloring.
All three strategies entail very similar ($<1\%$ of difference) times to
compute each iteration. Here, we select GrS and compare it to simple
pushing/pulling; see Figure~\ref{fig:bgc_analysis}.  Faster iterations are due
to fewer memory accesses as predicted in~\cref{sec:strategies}.
Next, we show that the strategies
differ in the number of iterations, see Table~\ref{tab:acc_iters}.
The largest iteration count (especially visible foe orc/ljn)
is due to FE. As predicted, this is because of
conflicts.  Both switching strategies reduce the iteration count.

\begin{figure}[!h]
\footnotesize
\sf
\centering
\subfloat{ 
\label{tab:acc_pr_perf}
\begin{tabular}{@{}l|ll@{}}
\toprule
$G$ & \textbf{Push} &    \textbf{+PA} \\\midrule
orc & 557.985 & 425.928  \\
pok & 103.907 & 87.577 \\
ljn & 240.943 & 145.475 \\
am  & 2.467   & 5.193 \\
rca & 5.422   & 13.705 \\\bottomrule
\end{tabular}
}
\hspace{1em}
\subfloat{ 
\label{tab:acc_iters}
\begin{tabular}{@{}l|llll@{}}
\toprule
$G$ & \textbf{Push} &    \textbf{+FE} & \textbf{+GS} & \textbf{+GrS}  \\\midrule
orc & 49 & 173 & 49 & 49 \\
pok & 49 & 48 & 49 & 47 \\
ljn & 49 & 334 & 49 & 49 \\
am  & 49 & 10 & 10 & 9 \\
rca & 49 & 5 & 5  & 5 \\
\bottomrule
\end{tabular}
}
\caption{(\cref{sec:acc_perf}) Acceleration strategy analysis (SM, Daint, XC30, $T=16$).
Time per iteration (ms) for PageRank (the left table).
Number of iterations to finish for BGC (the right table).
}
\label{fig:acc_perf_all}
\end{figure}

\macsubsection{Distributed-Memory Analysis}
\label{sec:dm-perf}


We also conduct a distributed-memory analysis.

\subsubsection{PageRank}
First, we use RMA for push- and pull-based PR.  The former uses remote
atomics (\texttt{MPI\_Accumulate}) to modify ranks. The latter read the
ranks with remote gets (\texttt{MPI\_Get}).
Next, we design PR with MP. Here, we use the collective
\texttt{MPI\_Alltoallv}~\cite{mpi3} to exchange the information on
the rank updates among processes. This variant is unusual as it
\emph{combines pushing and pulling}: each process contributes to the collective
by both providing a vector of rank updates (it pushes) and receiving updates
(it pulls).

\macb{Performance}
The performance outcomes (strong scaling) can be found in
Figure~\ref{fig:dm_analysis}.
MP consistently outperforms  
RMA (by $>$10x); pushing is the slowest. This may sound surprising
as MP comes with overheads due to buffer preparation. Contrarily to RMA, the
communicated updates must first be placed in designated send buffers.
Yet, 
the used
\texttt{MPI\_Accumulate} is implemented with costly underlying locking protocol.
Next, pulling suffers from
communication overheads as it fetches both the degree and the
rank of each neighbor of each vertex.

\macb{Memory Consumption}
RMA variants only use $\mathcal{O}(1)$ storage (per process) in
addition to the adjacency list. Contrarily, PR with MP 
may require up to $\mathcal{O}((n \hat{d})/P)$ storage (per process)
for send and receive buffers.

\subsubsection{Triangle Counting}
Similarly to PR, we develop push- and pull-based TC with RMA and with MP.
In pushing, we increase remote counters with an FAA. The MP-based TC 
uses messages to instruct which counters are augmented. To reduce communication
costs, updates are buffered until a given size is reached.

\macb{Performance}
The results are in Figure~\ref{fig:dm_analysis}. RMA variants
always outperform MP; pulling is always faster than pushing 
($<$1\% for orc and $\approx$25\% for ljn for $P=48$).  
This is different from PR as the counters in TC are
\emph{integer} and the utilized RMA library
offers fast path codes 
of remote atomic FAAs that access 64-bit integers. 
The MP variant is the
slowest because of the communication and buffering overheads.

\macb{Memory Consumption}
Both RMA schemes fetch $N(v)$ of each analyzed vertex $v$
to check for potential triangles. This is done with multiple
\texttt{MPI\_Get}s, with two extremes: a single get that fetches all the
neighbors, or one get per neighbor.  The former requires the largest amount of
additional memory ($\mathcal{O}(\hat{d})$ storage per process) but least
communication overheads. The latter is the opposite.

\macsubsection{Further Analyses}
\label{sec:more_analyses}

We now show that 
the relative differences between pushing and pulling do
not change significantly when varying the used machine.
We verify that PR comes with the most relevant difference; see
Table~\ref{tab:pr_analysis_machines}. Results vary most in denser
graphs (orc, pok, ljn); for example pushing outperforms pulling on Trivium while
the opposite is true on Dora. Contrarily, the results are similar for rca and am.
Thus, the overheads from branches, reads, and cache misses
(that are the highest in graphs with lowest $\overline{d}$) dominate performance.

\begin{table}[h]
\footnotesize
\sf
\centering
\begin{tabular}{@{}llllll@{}}
\toprule
                            Trivium: & orc    & pok    & ljn    & am     & rca    \\ \midrule
                            Push & 1426.966 & 191.340 & 373.134 & 6.199 & 16.818 \\
                            Pull & 1583.094 & 279.261 & 421.396 & 2.819 & 12.504 \\ 
                            Push+PA & 1289.123 & 190.541 & 400.634 & 8.549 & 52.068 \\ \midrule
                            Daint (XC40): & & & & & \\\midrule
                            Push & 499.463 & 123.784 & 248.602 & 5.744 & 7.753 \\
                            Pull & 456.532 & 86.812 & 206.604 & 2.828 & 5.810 \\ 
                            Push+PA & 378.548 & 78.883 & 128.255 & 6.157 & 14.102 \\ \bottomrule

                            \end{tabular}
\caption{(\cref{sec:more_analyses}) Time to compute one iteration in PR [ms] (SM setting with full parallelism (HT enabled); Trivium, $T=8$; Dora, XC40, $T=24$).}
\label{tab:pr_analysis_machines}
                            \end{table}


\macsubsection{Push-Pull Insights}

We finally summarize the most important insights on the push-pull performance
for the considered systems.


\macb{Shared-Memory Settings}
First, some algorithms are the fastest with pushing (SSSP-$\Delta$, BFS,
and PR for dense graphs) except for some data points (e.g., iteration~6
for orc in SSSP-$\Delta$). This contradicts the intuition that 
pulling comes with less overheads from atomics. Yet,
they either entail more reads that dominate performance (e.g.,
SSSP-$\Delta$) or use cache prefetchers less effectively by 
not accessing contiguous structures (e.g., PR). The results for PR+PA 
illustrate that atomics do not always dominate performance; this can happen if effects such
as cache misses become less dominant.
Second, SSSP-$\Delta$ on SM systems is surprisingly different
from the variant for the DM machines presented in the literature, where pulling
is faster~\cite{chakaravarthy2014scalable}. This is
because intra-node atomics are less costly than messages.
Next, HT accelerates each considered scheme, 
maintaining the relative differences between pushing and pulling.
Finally, several pulling schemes (in BGC and MST) are faster than their push counterparts.



\macb{Distributed-Memory Settings}
The choice of PR and TC illustrates that two algorithms with push and pull
variants having similar algorithm designs may come with substantially different
performance patterns.
Intuitively, RMA should ensure highest performance in both PR and TC as both
require the same \texttt{MPI\_Accumulate} remote atomic function.  Yet, the
different operand type results in different underlying implementations and thus
results.  With the setting considered in this work, 
RMA and MP ensured best performance for TC and PR, respectively.
%

\macsection{DISCUSSION}
\label{sec:discussion}

We now discuss various aspects of push and pull variants.


\macsubsection{Push-Pull: Linear Algebra}
\label{sec:la}

\goal{Introduce the LA graph view and motivate the PP analysis.}
Various graph algorithms can be expressed with linear algebra (LA) operations
such as matrix-vector (MV) multiplication. It enables a concise specification
by abstracting from details such as scheduling vertices
for processing in the next iteration~\cite{kepner2011graph}.
We now illustrate that it is possible to frame LA-based graph algorithms in
push and pull variants. 

\macb{Brief Recap}
\goal{Describe the required basics of LA-based graph schemes}
A crucial notion is the adjacency matrix of $G$ (denoted as
$\mathbf{A}$) that encodes $G$'s structure. The element in row~$i$ and
column~$j$ of $\mathbf{A}$ equals $1$ iff there is an edge from vertex $j$ to vertex $i$, 
and equals $0$ otherwise. For simplicity, we focus on unweighted
graphs, but our conclusions apply to the weighted case.

The graph algorithms that we consider can be cast as 
matrix-vector multiplications (MVs) $\mathbf{A} \otimes \mathbf{x}^{(k)}$,
where $\mathbf{x}^{(k)}$ is the algorithm state in iteration~$k$ and $\otimes$
is matrix-vector multiplication operator over an appropriate semiring.
The adjacency matrix $\mathbf{A}$ is generally sparse, while $\mathbf{x}^{(k)}$ 
may or may not be sparse depending on the computation.
For example, in PR, each $\mathbf{x}^{(k)}$ is dense, while in BFS, the
sparsity of $\mathbf{x}^{(k)}$ depends on the number of vertices in the $k$th frontier.
We refer to the case when the vector is dense as SpMV, and when the vector is sparse, SpMSpV.
The dichotomy between push and pull algorithm variants is mirrored by
the dichotomy between the Compressed Sparse Column (CSC) 
and Compressed Sparse Row (CSR) representations of $\mathbf{A}$.

A CSR representation stores each row of $\mathbf{A}$ contiguously.
The $i$th row of $\mathbf{A}$ contains all vertices with an edge to vertex~$i$.
Consequently, performing an SpMV in the CSR layout involves iterating over
each row and multiplying each nonzero element in the row by appropriate entries of the vector.
Thus, each entry of the output can be computed independently by a thread.
This scheme is equivalent to pulling updates for each vertex.
For SpMV, CSR (pulling) works extremely well, but for SpMSpV, it is not clear
how to efficiently exploit the sparsity of the vector $\mathbf{x}^{(k)}$.

A CSC representation stores each column of $\mathbf{A}$ contiguously.
The $i$th row of $\mathbf{A}$ contains all vertices with an edge from vertex~$i$.
Consequently, performing an SpMV in the CSC layout involves iterating over
each column and multiplying each nonzero element in the column by the same entry of the vector,
while accumulating to different elements of the output vector.
Here, atomics or a reduction tree are necessary to combine updates to each output vector element.
This scheme is equivalent to pushing updates from each vertex, as each thread is naturally
assigned a different column of $\mathbf{A}$ and nonzero entry of $\mathbf{x}^{(k)}$.
For SpMSpV, CSC (pushing) facilitates exploiting the sparsity of the vector by simply ignoring
columns of $\mathbf{A}$ that match up to zeros in $\mathbf{x}^{(k)}$.

\macsubsection{Push-Pull: Programming Models}

Push/pull differences depend on the programming model:

\macb{Threading/RMA}
The difference 
lies in the used atomics. An example is TC:
no atomics (pulling) and FAA (pushing).

\macb{MP (Point-to-Point Messages)}
In iterative
algorithms with fixed communication patterns (e.g., TC) pushing gives more speedup 
as pulling increases the message count.
In traversals, 
  pushing-pulling switching offers highest
  performance~\cite{beamer2013direction,chakaravarthy2014scalable}.

\macb{MP (Collectives)}
In collectives such as \texttt{MPI\_Alltoallv},
all processes both
push and pull the data, eliminating the distinction between these two.


\macsubsection{Push-Pull: Code Complexity}

Push and pull variants considered in this work come with similar code
complexity. Still, pull schemes can be more challenging in achieving high
performance. Consider the inner loop in PR where a thread iterates over $N(v)$
of a given $v$.  In pushing, updates are conducted simply with atomics.
Contrarily, in pulling, one must also fetch the degrees of neighbors. This is
similar for other pull variants and poses more challenges in making the code
fast.

\macsubsection{Push-Pull: Gather-Apply-Scatter}

Finally, we discuss the relationship between the push-pull dichotomy and the
well-know Gather-Apply-Scatter (GAS) abstraction~\cite{gonzalez2012powergraph}.
In GAS, one develops a graph algorithm by specifying the gather, apply, and
scatter functions. They run in parallel for each vertex $v$ and respectively:
bring some data from $v$'s neighbors, use it to modify $v$'s value, and write
the result to a data structure.
We now describe two algorithms designed with GAS (SSSP and
GC)~\cite{gonzalez2012powergraph} and show how to develop them with pushing or
pulling.

\macb{SSSP}
Here, each vertex $v$ is processed in parallel by selecting $v$'s incident
edge $e$ that offers a path to the selected root $s$ with the lowest distance.
If it is lower than the current distance from $v$ to $s$, the value is updated
accordingly and $N(s)$ are scheduled for processing in the next
iteration.
Now, push or pull can be applied when $v$ updates its distance to $s$.
In the former, a neighboring vertex that performed a relaxation in the previous
iteration updates its neighbors (pushes the changes) with new distances. In the
latter, each vertex scheduled for updates iterates over its neighbors (pulls
the updates) to perform a relaxation by itself.

\macb{GC}
Every vertex $v$ collects the set of colors on $N(v)$ to compute a
new unique color. Next, the new colors are scattered among $N(v)$. 
Any conflicting vertices are then scheduled for the color
recomputation in the next iteration.
This algorithm is a special case of BGC: each vertex constitutes a
separate partition (i.e., $\forall_{v \in V} \forall_{u \in N(v)} t[v] \neq
t[u]$).  Thus, the same approach can be incorporated.

\macsection{RELATED WORK}

\macb{Push and Pull Algorithm Variants}
Several graph algorithms that approach the pushing and pulling distinction have
been proposed. The bottom-up (pull) BFS was described by Suzumura et
al.~\cite{suzumura2011performance} while Beamer et
al.~\cite{beamer2013direction} introduced a direction-optimizing BFS that
switches between top-down (push) and bottom-up (pull) variants. Madduri et
al.~\cite{madduri2009faster} proposed several improvements to BC, one of which
inverts the direction of modifications in the backward traversal to eliminate
critical sections. Whang et al.~\cite{whang2015scalable} described pulling and
pushing in PR. Finally, Chakaravarthy et al.~\cite{chakaravarthy2014scalable}
inverts the direction of message exchanges in the distributed $\Delta$-Stepping
algorithm. All these schemes are solutions to single problems. We embrace and generalize
them in the push-pull analysis.
%

\sloppy
\macb{Pushing/Pulling in Graph Frameworks}
Various graph processing frameworks were introduced, for example
PBGL~\cite{Gregor05theparallel},
Pregel~\cite{Malewicz:2010:PSL:1807167.1807184},
GraphBLAS~\cite{mattson2014standards},
Galois~\cite{Kulkarni:2007:OPR:1250734.1250759},
HAMA~\cite{Seo:2010:HEM:1931470.1931872},
PowerGraph~\cite{gonzalez2012powergraph}, GraphLab~\cite{low2010graphlab}, and
Spark~\cite{Zaharia:2012:RDD:2228298.2228301}.
Some use pushing and pulling in certain ways, by: sending and receiving
messages (Pregel), using the GAS abstraction (PowerGraph), switching between
sparse and dense graph structures (Ligra~\cite{shun2013ligra}), switching the
direction of updates in a distributed environment
(Gemini~\cite{zhu2016gemini}), using pushing and pulling in 3D
task-partitioning~\cite{zhang2016exploring}, or pushing and pulling to/from
disk~\cite{wang2016hybrid}.  Yet, none of them comes with an analysis on the
push-pull dichotomy, focusing on the framework design.
Finally, Doekemeijer et al.~\cite{doekemeijer2014survey} list graph processing
frameworks that have push- or pull-based communication. 
Our theoretical analysis and performance observations can serve to help
better understand and improve graph processing frameworks.


\macb{Accelerating Strategies} The Grace
framework~\cite{prabhakaran2012managing} partitions the graph similarly to
Partition-Awareness, but its goal is to reduce caching overheads instead
of atomics in pushing.
Ligra uses a scheme similar to Generic-Switch as it switches between sparse and dense graph representations~\cite{shun2013ligra}.
Finally, Salihoglu et al.~\cite{salihoglu2014optimizing}
enhance Pregel-based systems with various schemes. Among others, similarly to Greedy-Switch, they
propose to switch from a Pregel-based distributed scheme to a sequential
algorithm variant.

\macb{Pushing/Pulling outside Graph Processing}
Borokhovich et al.~\cite{borokhovich2010tight} analyzed gossip algorithms in
network coding for information spreading using push, pull, and exchange
communication schemes. Swamy et al.~\cite{swamy2013asymptotically} designed an
asymptotically optimal push-pull method for multicasting over a random network.
Intel TBB uses a push-pull protocol in its flow graphs, biasing communication
to prevent polling and to reduce unnecessary retries~\cite{tbb-pp}.  An
analysis of push and pull in software engineering has also been
conducted~\cite{zhao2003model}.
None of these works addresses graph processing.

\macsection{CONCLUSION}

\goal{graph processing is important but still poses challenges}
Graph processing has become an important part of various CS research and
industry fields, including HPC, systems, networking, and architecture. Its
challenges, described by Lumsdaine et al.\ almost 10 years
ago~\cite{DBLP:journals/ppl/LumsdaineGHB07}, have still not been resolved and
accelerating graph computations remains an important goal that must be attained
for the ability to process the enormous amounts of data produced today.


\goal{We illustrate the PP view that enables new algorithms and speedups}
In this work, we accelerate graph algorithms by deriving the most advantageous
direction of graph updates out of the two options: \emph{pushing} the updates
from the private to the shared state, or \emph{pulling} the updates in the
opposite direction.  We illustrate in a detailed analysis that the \emph{Push-Pull (PP) dichotomy},
namely using either pushing or pulling, can be applied to various algorithms
such as triangle counting, minimum spanning tree computations, or graph
coloring. We provide detailed specifications, complexity analyses, and
performance data from hardware counters on which variant serves best each
algorithm and why pushing and pulling differ. These insights can be used to
improve various graph processing engines. 

Furthermore, we identify that pushing usually suffers from excessive amounts of
atomics/locks while pulling entails more memory reads/writes. We use 
generic strategies to limit the
amount of both, 
accelerating the processing of road networks,
citation graphs, social networks, and others.


Our analysis illustrates that the decision on using either
pushing or pulling is not limited to merely applying updates in PageRank or
sending messages in BFS, but is related to a wide class of algorithms,
strategies, graph abstractions, and programming models.
Our PP dichotomy can easily be generalized to other concepts related to graph
processing, for example vectorization~\cite{besta2017slimsell}.

\section*{Acknowledgments}

We thank Hussein Harake, Colin McMurtrie, and the whole CSCS team granting access to the Greina, Piz Dora, and Daint machines, and
for their excellent technical support. 

\bibliographystyle{ACM-Reference-Format}
\bibliography{references}


\begin{thebibliography}{00}


\ifx \showCODEN    \undefined \def \showCODEN     #1{\unskip}     \fi
\ifx \showDOI      \undefined \def \showDOI       #1{#1}\fi
\ifx \showISBNx    \undefined \def \showISBNx     #1{\unskip}     \fi
\ifx \showISBNxiii \undefined \def \showISBNxiii  #1{\unskip}     \fi
\ifx \showISSN     \undefined \def \showISSN      #1{\unskip}     \fi
\ifx \showLCCN     \undefined \def \showLCCN      #1{\unskip}     \fi
\ifx \shownote     \undefined \def \shownote      #1{#1}          \fi
\ifx \showarticletitle \undefined \def \showarticletitle #1{#1}   \fi
\ifx \showURL      \undefined \def \showURL       {\relax}        \fi
\providecommand\bibfield[2]{#2}
\providecommand\bibinfo[2]{#2}
\providecommand\natexlab[1]{#1}
\providecommand\showeprint[2][]{arXiv:#2}

\bibitem[\protect\citeauthoryear{Awerbuch and Shiloach}{Awerbuch and
  Shiloach}{1987}]%
        {awerbuch1987new}
\bibfield{author}{\bibinfo{person}{Baruch Awerbuch} {and}
  \bibinfo{person}{Yossi Shiloach}.} \bibinfo{year}{1987}\natexlab{}.
\newblock \showarticletitle{{New connectivity and MSF algorithms for
  shuffle-exchange network and PRAM}}.
\newblock \bibinfo{journal}{{\em IEEE Trans. on Comp.\/}} \bibinfo{volume}{36},
  \bibinfo{number}{10} (\bibinfo{year}{1987}), \bibinfo{pages}{1258--1263}.
\newblock


\bibitem[\protect\citeauthoryear{Bader et~al\mbox{.}}{Bader
  et~al\mbox{.}}{2007}]%
        {bader2007approximating}
\bibfield{author}{\bibinfo{person}{David~A Bader} {et~al\mbox{.}}}
  \bibinfo{year}{2007}\natexlab{}.
\newblock \showarticletitle{{Approximating betweenness centrality}}.
\newblock In \bibinfo{booktitle}{{\em Algorithms and Models for the
  Web-Graph}}. \bibinfo{publisher}{Springer}, \bibinfo{pages}{124--137}.
\newblock


\bibitem[\protect\citeauthoryear{Bader and Cong}{Bader and Cong}{2004}]%
        {bader2004fast}
\bibfield{author}{\bibinfo{person}{David~A Bader} {and}
  \bibinfo{person}{Guojing Cong}.} \bibinfo{year}{2004}\natexlab{}.
\newblock \showarticletitle{Fast shared-memory algorithms for computing the
  minimum spanning forest of sparse graphs}. In \bibinfo{booktitle}{{\em Par.
  and Dist. Proc. Symp. (IPDPS)}}. IEEE, \bibinfo{pages}{39}.
\newblock


\bibitem[\protect\citeauthoryear{Beamer, Asanovi{\'c}, and Patterson}{Beamer
  et~al\mbox{.}}{2013}]%
        {beamer2013direction}
\bibfield{author}{\bibinfo{person}{Scott Beamer}, \bibinfo{person}{Krste
  Asanovi{\'c}}, {and} \bibinfo{person}{David Patterson}.}
  \bibinfo{year}{2013}\natexlab{}.
\newblock \showarticletitle{{Direction-optimizing breadth-first search}}.
\newblock \bibinfo{journal}{{\em Scientific Programming\/}}
  \bibinfo{volume}{21}, \bibinfo{number}{3-4} (\bibinfo{year}{2013}),
  \bibinfo{pages}{137--148}.
\newblock


\bibitem[\protect\citeauthoryear{Beamer, Asanovi{\'c}, and Patterson}{Beamer
  et~al\mbox{.}}{2015}]%
        {beamer2015gail}
\bibfield{author}{\bibinfo{person}{Scott Beamer}, \bibinfo{person}{Krste
  Asanovi{\'c}}, {and} \bibinfo{person}{David Patterson}.}
  \bibinfo{year}{2015}\natexlab{}.
\newblock \showarticletitle{{GAIL: the graph algorithm iron law}}. In
  \bibinfo{booktitle}{{\em Workshop on Ir. App.: Arch. and Alg.}}
  \bibinfo{pages}{13}.
\newblock


\bibitem[\protect\citeauthoryear{Besta and Hoefler}{Besta and Hoefler}{2014a}]%
        {besta2014fault}
\bibfield{author}{\bibinfo{person}{Maciej Besta} {and} \bibinfo{person}{Torsten
  Hoefler}.} \bibinfo{year}{2014}\natexlab{a}.
\newblock \showarticletitle{Fault tolerance for remote memory access
  programming models}. In \bibinfo{booktitle}{{\em Proceedings of the 23rd
  international symposium on High-performance parallel and distributed
  computing}}. \bibinfo{pages}{37--48}.
\newblock


\bibitem[\protect\citeauthoryear{Besta and Hoefler}{Besta and Hoefler}{2014b}]%
        {besta2014slim}
\bibfield{author}{\bibinfo{person}{Maciej Besta} {and} \bibinfo{person}{Torsten
  Hoefler}.} \bibinfo{year}{2014}\natexlab{b}.
\newblock \showarticletitle{Slim fly: A cost effective low-diameter network
  topology}. In \bibinfo{booktitle}{{\em SC'14: Proceedings of the
  International Conference for High Performance Computing, Networking, Storage
  and Analysis}}. IEEE, \bibinfo{pages}{348--359}.
\newblock


\bibitem[\protect\citeauthoryear{Besta and Hoefler}{Besta and Hoefler}{2015a}]%
        {besta2015accelerating}
\bibfield{author}{\bibinfo{person}{Maciej Besta} {and} \bibinfo{person}{Torsten
  Hoefler}.} \bibinfo{year}{2015}\natexlab{a}.
\newblock \showarticletitle{Accelerating irregular computations with hardware
  transactional memory and active messages}. In \bibinfo{booktitle}{{\em
  Proceedings of the 24th International Symposium on High-Performance Parallel
  and Distributed Computing}}. \bibinfo{pages}{161--172}.
\newblock


\bibitem[\protect\citeauthoryear{Besta and Hoefler}{Besta and Hoefler}{2015b}]%
        {besta2015active}
\bibfield{author}{\bibinfo{person}{Maciej Besta} {and} \bibinfo{person}{Torsten
  Hoefler}.} \bibinfo{year}{2015}\natexlab{b}.
\newblock \showarticletitle{Active access: A mechanism for high-performance
  distributed data-centric computations}. In \bibinfo{booktitle}{{\em
  Proceedings of the 29th ACM on International Conference on Supercomputing}}.
  \bibinfo{pages}{155--164}.
\newblock


\bibitem[\protect\citeauthoryear{Besta, Marending, Solomonik, and
  Hoefler}{Besta et~al\mbox{.}}{2017}]%
        {besta2017slimsell}
\bibfield{author}{\bibinfo{person}{Maciej Besta}, \bibinfo{person}{Florian
  Marending}, \bibinfo{person}{Edgar Solomonik}, {and} \bibinfo{person}{Torsten
  Hoefler}.} \bibinfo{year}{2017}\natexlab{}.
\newblock \showarticletitle{Slimsell: A vectorizable graph representation for
  breadth-first search}. In \bibinfo{booktitle}{{\em 2017 IEEE International
  Parallel and Distributed Processing Symposium (IPDPS)}}. IEEE,
  \bibinfo{pages}{32--41}.
\newblock


\bibitem[\protect\citeauthoryear{Boman et~al\mbox{.}}{Boman
  et~al\mbox{.}}{2005}]%
        {boman2005scalable}
\bibfield{author}{\bibinfo{person}{Erik~G Boman} {et~al\mbox{.}}}
  \bibinfo{year}{2005}\natexlab{}.
\newblock \showarticletitle{A scalable parallel graph coloring algorithm for
  distributed memory computers}.
\newblock In \bibinfo{booktitle}{{\em Euro-Par}}. \bibinfo{pages}{241--251}.
\newblock


\bibitem[\protect\citeauthoryear{Borokhovich et~al\mbox{.}}{Borokhovich
  et~al\mbox{.}}{2010}]%
        {borokhovich2010tight}
\bibfield{author}{\bibinfo{person}{Michael Borokhovich} {et~al\mbox{.}}}
  \bibinfo{year}{2010}\natexlab{}.
\newblock \showarticletitle{{Tight bounds for algebraic gossip on graphs}}. In
  \bibinfo{booktitle}{{\em Inf. Theory Proc. (ISIT), IEEE Intl. Symp. on}}.
  \bibinfo{pages}{1758--1762}.
\newblock


\bibitem[\protect\citeauthoryear{Boruvka}{Boruvka}{1926}]%
        {boruuvka1926jistem}
\bibfield{author}{\bibinfo{person}{Otakar Boruvka}.}
  \bibinfo{year}{1926}\natexlab{}.
\newblock \showarticletitle{O jist{\'e}m probl{\'e}mu minim{\'a}ln{\'\i}m}.
\newblock  (\bibinfo{year}{1926}).
\newblock


\bibitem[\protect\citeauthoryear{Brandes}{Brandes}{2001}]%
        {brandes2001faster}
\bibfield{author}{\bibinfo{person}{Ulrik Brandes}.}
  \bibinfo{year}{2001}\natexlab{}.
\newblock \showarticletitle{A faster algorithm for betweenness centrality}.
\newblock \bibinfo{journal}{{\em J. of Math. Sociology\/}}
  \bibinfo{volume}{25}, \bibinfo{number}{2} (\bibinfo{year}{2001}),
  \bibinfo{pages}{163--177}.
\newblock


\bibitem[\protect\citeauthoryear{Brin and Page}{Brin and Page}{1998}]%
        {Brin:1998:ALH:297805.297827}
\bibfield{author}{\bibinfo{person}{Sergey Brin} {and} \bibinfo{person}{Lawrence
  Page}.} \bibinfo{year}{1998}\natexlab{}.
\newblock \showarticletitle{{The anatomy of a large-scale hypertextual Web
  search engine}}. In \bibinfo{booktitle}{{\em Proc. of Intl. Conf. on World
  Wide Web}} {\em (\bibinfo{series}{WWW7})}. \bibinfo{pages}{107--117}.
\newblock


\bibitem[\protect\citeauthoryear{Catalyurek and Aykanat}{Catalyurek and
  Aykanat}{2001}]%
        {Catalyurek:2001:FHM:645609.663255}
\bibfield{author}{\bibinfo{person}{Umit Catalyurek} {and}
  \bibinfo{person}{Cevdet Aykanat}.} \bibinfo{year}{2001}\natexlab{}.
\newblock \showarticletitle{{A Fine-Grain Hypergraph Model for 2D Decomposition
  of Sparse Matrices}}. In \bibinfo{booktitle}{{\em Proc. of the Intl. Par.
  \&Amp; Dist. Proc. Symp.}} {\em (\bibinfo{series}{IPDPS '01})}.
  \bibinfo{pages}{118--}.
\newblock
\showISBNx{0-7695-0990-8}
\showURL{%
\url{http://dl.acm.org/citation.cfm?id=645609.663255}}


\bibitem[\protect\citeauthoryear{Chakaravarthy et~al\mbox{.}}{Chakaravarthy
  et~al\mbox{.}}{2014}]%
        {chakaravarthy2014scalable}
\bibfield{author}{\bibinfo{person}{Venkatesan~T Chakaravarthy} {et~al\mbox{.}}}
  \bibinfo{year}{2014}\natexlab{}.
\newblock \showarticletitle{{Scalable single source shortest path algorithms
  for massively parallel systems}}. In \bibinfo{booktitle}{{\em Par. and Dist.
  Proc. Symp., IEEE Intl.}} \bibinfo{pages}{889--901}.
\newblock


\bibitem[\protect\citeauthoryear{Cormen, Stein, Rivest, and Leiserson}{Cormen
  et~al\mbox{.}}{2001}]%
        {Cormen:2001:IA:580470}
\bibfield{author}{\bibinfo{person}{Thomas~H. Cormen}, \bibinfo{person}{Clifford
  Stein}, \bibinfo{person}{Ronald~L. Rivest}, {and} \bibinfo{person}{Charles~E.
  Leiserson}.} \bibinfo{year}{2001}\natexlab{}.
\newblock \bibinfo{booktitle}{{\em {Introduction to Algorithms}\/}
  (\bibinfo{edition}{2nd} ed.)}.
\newblock \bibinfo{publisher}{McGraw-Hill Higher Education}.
\newblock
\showISBNx{0070131511}


\bibitem[\protect\citeauthoryear{Csardi and Nepusz}{Csardi and Nepusz}{2006}]%
        {csardi2006igraph}
\bibfield{author}{\bibinfo{person}{Gabor Csardi} {and} \bibinfo{person}{Tamas
  Nepusz}.} \bibinfo{year}{2006}\natexlab{}.
\newblock \showarticletitle{{The igraph software package for complex network
  research}}.
\newblock \bibinfo{journal}{{\em InterJournal, Complex Systems\/}}
  \bibinfo{volume}{1695}, \bibinfo{number}{5} (\bibinfo{year}{2006}),
  \bibinfo{pages}{1--9}.
\newblock


\bibitem[\protect\citeauthoryear{Doekemeijer and Varbanescu}{Doekemeijer and
  Varbanescu}{2014}]%
        {doekemeijer2014survey}
\bibfield{author}{\bibinfo{person}{Niels Doekemeijer} {and}
  \bibinfo{person}{Ana~Lucia Varbanescu}.} \bibinfo{year}{2014}\natexlab{}.
\newblock \showarticletitle{A survey of parallel graph processing frameworks}.
\newblock \bibinfo{journal}{{\em Delft University of Technology\/}}
  (\bibinfo{year}{2014}).
\newblock


\bibitem[\protect\citeauthoryear{Erd{\H{o}}s and R{\'e}nyi}{Erd{\H{o}}s and
  R{\'e}nyi}{1976}]%
        {erdHos1976evolution}
\bibfield{author}{\bibinfo{person}{P Erd{\H{o}}s} {and} \bibinfo{person}{A
  R{\'e}nyi}.} \bibinfo{year}{1976}\natexlab{}.
\newblock \showarticletitle{On the evolution of random graphs}.
\newblock \bibinfo{journal}{{\em Selected Papers of Alfr{\'e}d R{\'e}nyi\/}}
  \bibinfo{volume}{2} (\bibinfo{year}{1976}), \bibinfo{pages}{482--525}.
\newblock


\bibitem[\protect\citeauthoryear{Fortune and Wyllie}{Fortune and
  Wyllie}{1978}]%
        {fortune1978parallelism}
\bibfield{author}{\bibinfo{person}{Steven Fortune} {and} \bibinfo{person}{James
  Wyllie}.} \bibinfo{year}{1978}\natexlab{}.
\newblock \showarticletitle{{Parallelism in random access machines}}. In
  \bibinfo{booktitle}{{\em Proc. of ACM Symp. on Theory of Comp.}}
  \bibinfo{pages}{114--118}.
\newblock


\bibitem[\protect\citeauthoryear{Gazit et~al\mbox{.}}{Gazit
  et~al\mbox{.}}{1988a}]%
        {gazit1988improved}
\bibfield{author}{\bibinfo{person}{Hillel Gazit} {et~al\mbox{.}}}
  \bibinfo{year}{1988}\natexlab{a}.
\newblock \showarticletitle{{An improved parallel algorithm that computes the
  BFS numbering of a directed graph}}.
\newblock \bibinfo{journal}{{\em Inf. Proc. Let.\/}} \bibinfo{volume}{28},
  \bibinfo{number}{2} (\bibinfo{year}{1988}), \bibinfo{pages}{61--65}.
\newblock


\bibitem[\protect\citeauthoryear{Gazit et~al\mbox{.}}{Gazit
  et~al\mbox{.}}{1988b}]%
        {gazit1988optimal}
\bibfield{author}{\bibinfo{person}{Hillel Gazit} {et~al\mbox{.}}}
  \bibinfo{year}{1988}\natexlab{b}.
\newblock \showarticletitle{Optimal tree contraction in the {EREW} model}.
\newblock In \bibinfo{booktitle}{{\em Concurrent Computations}}.
  \bibinfo{publisher}{Springer}, \bibinfo{pages}{139--156}.
\newblock


\bibitem[\protect\citeauthoryear{Gerstenberger, Besta, and
  Hoefler}{Gerstenberger et~al\mbox{.}}{2013}]%
        {gerstenberger2013enabling}
\bibfield{author}{\bibinfo{person}{R. Gerstenberger}, \bibinfo{person}{M.
  Besta}, {and} \bibinfo{person}{T. Hoefler}.} \bibinfo{year}{2013}\natexlab{}.
\newblock \showarticletitle{{Enabling Highly-scalable Remote Memory Access
  Programming with MPI-3 One Sided}}. In \bibinfo{booktitle}{{\em Proc. of the
  ACM/IEEE Supercomputing}} {\em (\bibinfo{series}{SC '13})}. Article
  \bibinfo{articleno}{53}, \bibinfo{numpages}{12}~pages.
\newblock
\showISBNx{978-1-4503-2378-9}


\bibitem[\protect\citeauthoryear{Goel and Munagala}{Goel and Munagala}{2012}]%
        {goel2012complexity}
\bibfield{author}{\bibinfo{person}{Ashish Goel} {and} \bibinfo{person}{Kamesh
  Munagala}.} \bibinfo{year}{2012}\natexlab{}.
\newblock \showarticletitle{Complexity measures for map-reduce, and comparison
  to parallel computing}.
\newblock \bibinfo{journal}{{\em arXiv preprint arXiv:1211.6526\/}}
  (\bibinfo{year}{2012}).
\newblock


\bibitem[\protect\citeauthoryear{Gonzalez et~al\mbox{.}}{Gonzalez
  et~al\mbox{.}}{2012}]%
        {gonzalez2012powergraph}
\bibfield{author}{\bibinfo{person}{Joseph~E Gonzalez} {et~al\mbox{.}}}
  \bibinfo{year}{2012}\natexlab{}.
\newblock \showarticletitle{{PowerGraph: Distributed Graph-Parallel Computation
  on Natural Graphs.}}. In \bibinfo{booktitle}{{\em OSDI}},
  Vol.~\bibinfo{volume}{12}. \bibinfo{pages}{2}.
\newblock


\bibitem[\protect\citeauthoryear{Green et~al\mbox{.}}{Green
  et~al\mbox{.}}{2014}]%
        {green2014branch}
\bibfield{author}{\bibinfo{person}{Oded Green} {et~al\mbox{.}}}
  \bibinfo{year}{2014}\natexlab{}.
\newblock \showarticletitle{{Branch-Avoiding Graph Algorithms}}.
\newblock \bibinfo{journal}{{\em arXiv:1411.1460\/}} (\bibinfo{year}{2014}).
\newblock


\bibitem[\protect\citeauthoryear{Gregor and Lumsdaine}{Gregor and
  Lumsdaine}{2005}]%
        {Gregor05theparallel}
\bibfield{author}{\bibinfo{person}{Douglas Gregor} {and}
  \bibinfo{person}{Andrew Lumsdaine}.} \bibinfo{year}{2005}\natexlab{}.
\newblock \showarticletitle{{The parallel BGL: A generic library for
  distributed graph computations}}.
\newblock \bibinfo{journal}{{\em Par. Obj.-Or. Scientific Comp. (POOSC)\/}}
  (\bibinfo{year}{2005}), \bibinfo{pages}{2}.
\newblock


\bibitem[\protect\citeauthoryear{Harris}{Harris}{1994}]%
        {harris1994survey}
\bibfield{author}{\bibinfo{person}{Tim~J Harris}.}
  \bibinfo{year}{1994}\natexlab{}.
\newblock \showarticletitle{{A survey of PRAM simulation techniques}}.
\newblock \bibinfo{journal}{{\em ACM Comp. Surv. (CSUR)\/}}
  \bibinfo{volume}{26}, \bibinfo{number}{2} (\bibinfo{year}{1994}),
  \bibinfo{pages}{187--206}.
\newblock


\bibitem[\protect\citeauthoryear{{Intel, Inc.}}{{Intel, Inc.}}{2015}]%
        {intel64and}
\bibfield{author}{\bibinfo{person}{{Intel, Inc.}}}
  \bibinfo{year}{2015}\natexlab{}.
\newblock \bibinfo{title}{{64 and IA-32 Architectures Software Developer’s
  Manual}}.
\newblock   (\bibinfo{year}{2015}).
\newblock


\bibitem[\protect\citeauthoryear{Kepner and Gilbert}{Kepner and
  Gilbert}{2011}]%
        {kepner2011graph}
\bibfield{author}{\bibinfo{person}{Jeremy Kepner} {and} \bibinfo{person}{John
  Gilbert}.} \bibinfo{year}{2011}\natexlab{}.
\newblock \bibinfo{booktitle}{{\em Graph algorithms in the language of linear
  algebra}}. Vol.~\bibinfo{volume}{22}.
\newblock \bibinfo{publisher}{SIAM}.
\newblock


\bibitem[\protect\citeauthoryear{Kim et~al\mbox{.}}{Kim et~al\mbox{.}}{2008}]%
        {dally08}
\bibfield{author}{\bibinfo{person}{John Kim} {et~al\mbox{.}}}
  \bibinfo{year}{2008}\natexlab{}.
\newblock \showarticletitle{{Technology-Driven, Highly-Scalable Dragonfly
  Topology}}. In \bibinfo{booktitle}{{\em Ann. Intl. Symp. on Comp. Arch.}}
  {\em (\bibinfo{series}{ISCA '08})}. \bibinfo{pages}{77--88}.
\newblock
\showISBNx{978-0-7695-3174-8}
\showDOI{%
\url{https://doi.org/10.1109/ISCA.2008.19}}


\bibitem[\protect\citeauthoryear{Kulkarni et~al\mbox{.}}{Kulkarni
  et~al\mbox{.}}{2007}]%
        {Kulkarni:2007:OPR:1250734.1250759}
\bibfield{author}{\bibinfo{person}{Milind Kulkarni} {et~al\mbox{.}}}
  \bibinfo{year}{2007}\natexlab{}.
\newblock \showarticletitle{{Optimistic parallelism requires abstractions}}. In
  \bibinfo{booktitle}{{\em ACM SIGPLAN Conf. on Prog. Lang. Des. and Impl.}}
  {\em (\bibinfo{series}{PLDI '07})}. \bibinfo{pages}{211--222}.
\newblock
\showISBNx{978-1-59593-633-2}
\showDOI{%
\url{https://doi.org/10.1145/1250734.1250759}}


\bibitem[\protect\citeauthoryear{Leiserson and Schardl}{Leiserson and
  Schardl}{2010}]%
        {leiserson2010work}
\bibfield{author}{\bibinfo{person}{Charles~E Leiserson} {and}
  \bibinfo{person}{Tao~B Schardl}.} \bibinfo{year}{2010}\natexlab{}.
\newblock \showarticletitle{A work-efficient parallel breadth-first search
  algorithm (or how to cope with the nondeterminism of reducers)}. In
  \bibinfo{booktitle}{{\em Proc. of ACM Symp. on Par. in Alg. and Arch.}}
  \bibinfo{pages}{303--314}.
\newblock


\bibitem[\protect\citeauthoryear{Leskovec et~al\mbox{.}}{Leskovec
  et~al\mbox{.}}{2010}]%
        {leskovec2010kronecker}
\bibfield{author}{\bibinfo{person}{Jure Leskovec} {et~al\mbox{.}}}
  \bibinfo{year}{2010}\natexlab{}.
\newblock \showarticletitle{{Kronecker graphs: An approach to modeling
  networks}}.
\newblock \bibinfo{journal}{{\em J. of Machine Learning Research\/}}
  \bibinfo{volume}{11}, \bibinfo{number}{Feb} (\bibinfo{year}{2010}),
  \bibinfo{pages}{985--1042}.
\newblock


\bibitem[\protect\citeauthoryear{Low et~al\mbox{.}}{Low et~al\mbox{.}}{2010}]%
        {low2010graphlab}
\bibfield{author}{\bibinfo{person}{Yucheng Low} {et~al\mbox{.}}}
  \bibinfo{year}{2010}\natexlab{}.
\newblock \showarticletitle{{Graphlab: A new framework for parallel machine
  learning}}.
\newblock \bibinfo{journal}{{\em preprint arXiv:1006.4990\/}}
  (\bibinfo{year}{2010}).
\newblock


\bibitem[\protect\citeauthoryear{Lumsdaine, Gregor, Hendrickson, and
  Berry}{Lumsdaine et~al\mbox{.}}{2007}]%
        {DBLP:journals/ppl/LumsdaineGHB07}
\bibfield{author}{\bibinfo{person}{Andrew Lumsdaine}, \bibinfo{person}{Douglas
  Gregor}, \bibinfo{person}{Bruce Hendrickson}, {and}
  \bibinfo{person}{Jonathan~W. Berry}.} \bibinfo{year}{2007}\natexlab{}.
\newblock \showarticletitle{{Challenges in Parallel Graph Processing}}.
\newblock \bibinfo{journal}{{\em Par. Proc. Let.\/}} \bibinfo{volume}{17},
  \bibinfo{number}{1} (\bibinfo{year}{2007}), \bibinfo{pages}{5--20}.
\newblock


\bibitem[\protect\citeauthoryear{Madduri et~al\mbox{.}}{Madduri
  et~al\mbox{.}}{2009}]%
        {madduri2009faster}
\bibfield{author}{\bibinfo{person}{Kamesh Madduri} {et~al\mbox{.}}}
  \bibinfo{year}{2009}\natexlab{}.
\newblock \showarticletitle{{A faster parallel algorithm and efficient
  multithreaded implementations for evaluating betweenness centrality on
  massive datasets}}. In \bibinfo{booktitle}{{\em Par. \& Dist. Proc. (IPDPS),
  IEEE Intl. Symp. on}}. \bibinfo{pages}{1--8}.
\newblock


\bibitem[\protect\citeauthoryear{Malewicz et~al\mbox{.}}{Malewicz
  et~al\mbox{.}}{2010}]%
        {Malewicz:2010:PSL:1807167.1807184}
\bibfield{author}{\bibinfo{person}{Grzegorz Malewicz} {et~al\mbox{.}}}
  \bibinfo{year}{2010}\natexlab{}.
\newblock \showarticletitle{{Pregel: a system for large-scale graph
  processing}}. In \bibinfo{booktitle}{{\em ACM SIGMOD Intl. Conf. on Manag. of
  Data}} {\em (\bibinfo{series}{SIGMOD '10})}. \bibinfo{pages}{135--146}.
\newblock
\showISBNx{978-1-4503-0032-2}
\showDOI{%
\url{https://doi.org/10.1145/1807167.1807184}}


\bibitem[\protect\citeauthoryear{Mattson et~al\mbox{.}}{Mattson
  et~al\mbox{.}}{2014}]%
        {mattson2014standards}
\bibfield{author}{\bibinfo{person}{Tim Mattson} {et~al\mbox{.}}}
  \bibinfo{year}{2014}\natexlab{}.
\newblock \showarticletitle{{Standards for graph algorithm primitives}}.
\newblock \bibinfo{journal}{{\em arXiv preprint arXiv:1408.0393\/}}
  (\bibinfo{year}{2014}).
\newblock


\bibitem[\protect\citeauthoryear{Meyer and Sanders}{Meyer and Sanders}{2003}]%
        {meyer2003delta}
\bibfield{author}{\bibinfo{person}{Ulrich Meyer} {and} \bibinfo{person}{Peter
  Sanders}.} \bibinfo{year}{2003}\natexlab{}.
\newblock \showarticletitle{{$\Delta$-stepping: a parallelizable shortest path
  algorithm}}.
\newblock \bibinfo{journal}{{\em Journal of Algorithms\/}}
  \bibinfo{volume}{49}, \bibinfo{number}{1} (\bibinfo{year}{2003}),
  \bibinfo{pages}{114--152}.
\newblock


\bibitem[\protect\citeauthoryear{{MPI Forum}}{{MPI Forum}}{2012}]%
        {mpi3}
\bibfield{author}{\bibinfo{person}{{MPI Forum}}.}
  \bibinfo{year}{2012}\natexlab{}.
\newblock \bibinfo{title}{{\textsf{MPI}: A Message-Passing Interface Standard.
  Version 3}}.
\newblock   (\bibinfo{year}{2012}).
\newblock


\bibitem[\protect\citeauthoryear{Murphy et~al\mbox{.}}{Murphy
  et~al\mbox{.}}{2010}]%
        {murphy2010introducing}
\bibfield{author}{\bibinfo{person}{Richard~C Murphy} {et~al\mbox{.}}}
  \bibinfo{year}{2010}\natexlab{}.
\newblock \showarticletitle{Introducing the graph 500}.
\newblock \bibinfo{journal}{{\em Cray User’s Group (CUG)\/}}
  (\bibinfo{year}{2010}).
\newblock


\bibitem[\protect\citeauthoryear{Prabhakaran et~al\mbox{.}}{Prabhakaran
  et~al\mbox{.}}{2012}]%
        {prabhakaran2012managing}
\bibfield{author}{\bibinfo{person}{Vijayan Prabhakaran} {et~al\mbox{.}}}
  \bibinfo{year}{2012}\natexlab{}.
\newblock \showarticletitle{Managing Large Graphs on Multi-Cores with Graph
  Awareness.}. In \bibinfo{booktitle}{{\em USENIX Annual Technical
  Conference}}, Vol.~\bibinfo{volume}{12}.
\newblock


\bibitem[\protect\citeauthoryear{Prountzos and Pingali}{Prountzos and
  Pingali}{2013}]%
        {prountzos2013betweenness}
\bibfield{author}{\bibinfo{person}{Dimitrios Prountzos} {and}
  \bibinfo{person}{Keshav Pingali}.} \bibinfo{year}{2013}\natexlab{}.
\newblock \showarticletitle{{Betweenness centrality: algorithms and
  implementations}}. In \bibinfo{booktitle}{{\em ACM SIGPLAN Notices}},
  Vol.~\bibinfo{volume}{48}. ACM, \bibinfo{pages}{35--46}.
\newblock


\bibitem[\protect\citeauthoryear{Salihoglu and Widom}{Salihoglu and
  Widom}{2014}]%
        {salihoglu2014optimizing}
\bibfield{author}{\bibinfo{person}{Semih Salihoglu} {and}
  \bibinfo{person}{Jennifer Widom}.} \bibinfo{year}{2014}\natexlab{}.
\newblock \showarticletitle{{Optimizing graph algorithms on Pregel-like
  systems}}.
\newblock \bibinfo{journal}{{\em Proceedings of the VLDB Endowment\/}}
  \bibinfo{volume}{7}, \bibinfo{number}{7} (\bibinfo{year}{2014}),
  \bibinfo{pages}{577--588}.
\newblock


\bibitem[\protect\citeauthoryear{Satish et~al\mbox{.}}{Satish
  et~al\mbox{.}}{2014}]%
        {satish2014navigating}
\bibfield{author}{\bibinfo{person}{Nadathur Satish} {et~al\mbox{.}}}
  \bibinfo{year}{2014}\natexlab{}.
\newblock \showarticletitle{Navigating the maze of graph analytics frameworks
  using massive graph datasets}. In \bibinfo{booktitle}{{\em ACM SIGMOD Intl.
  Conf. on Man. Data}}. \bibinfo{pages}{979--990}.
\newblock


\bibitem[\protect\citeauthoryear{Schank}{Schank}{2007}]%
        {schank2007algorithmic}
\bibfield{author}{\bibinfo{person}{Thomas Schank}.}
  \bibinfo{year}{2007}\natexlab{}.
\newblock {\em \bibinfo{title}{Algorithmic aspects of triangle-based network
  analysis}}.
\newblock \bibinfo{thesistype}{Ph.D. Dissertation}. \bibinfo{school}{University
  Karlsruhe}.
\newblock


\bibitem[\protect\citeauthoryear{Schweizer, Besta, and Hoefler}{Schweizer
  et~al\mbox{.}}{2015}]%
        {schweizer2015evaluating}
\bibfield{author}{\bibinfo{person}{Hermann Schweizer}, \bibinfo{person}{Maciej
  Besta}, {and} \bibinfo{person}{Torsten Hoefler}.}
  \bibinfo{year}{2015}\natexlab{}.
\newblock \showarticletitle{Evaluating the cost of atomic operations on modern
  architectures}. In \bibinfo{booktitle}{{\em 2015 International Conference on
  Parallel Architecture and Compilation (PACT)}}. IEEE,
  \bibinfo{pages}{445--456}.
\newblock


\bibitem[\protect\citeauthoryear{Seo et~al\mbox{.}}{Seo et~al\mbox{.}}{2010}]%
        {Seo:2010:HEM:1931470.1931872}
\bibfield{author}{\bibinfo{person}{Sangwon Seo} {et~al\mbox{.}}}
  \bibinfo{year}{2010}\natexlab{}.
\newblock \showarticletitle{{HAMA: An Efficient Matrix Computation with the
  MapReduce Framework}}. In \bibinfo{booktitle}{{\em Intl. Conf. on Cloud Comp.
  Tech. and Science}} {\em (\bibinfo{series}{CLOUDCOM'10})}.
  \bibinfo{pages}{721--726}.
\newblock
\showISBNx{978-0-7695-4302-4}
\showDOI{%
\url{https://doi.org/10.1109/CloudCom.2010.17}}


\bibitem[\protect\citeauthoryear{Shun and Blelloch}{Shun and Blelloch}{2013}]%
        {shun2013ligra}
\bibfield{author}{\bibinfo{person}{Julian Shun} {and} \bibinfo{person}{Guy~E
  Blelloch}.} \bibinfo{year}{2013}\natexlab{}.
\newblock \showarticletitle{{Ligra: a lightweight graph processing framework
  for shared memory}}. In \bibinfo{booktitle}{{\em ACM SIGPLAN Notices}},
  Vol.~\bibinfo{volume}{48}. \bibinfo{pages}{135--146}.
\newblock


\bibitem[\protect\citeauthoryear{Shun and Tangwongsan}{Shun and
  Tangwongsan}{2015}]%
        {7113280}
\bibfield{author}{\bibinfo{person}{J. Shun} {and} \bibinfo{person}{K.
  Tangwongsan}.} \bibinfo{year}{2015}\natexlab{}.
\newblock \showarticletitle{Multicore triangle computations without tuning}. In
  \bibinfo{booktitle}{{\em 2015 IEEE 31st Intl. Conf. on Data Engineering}}.
  \bibinfo{pages}{149--160}.
\newblock
\showISSN{1063-6382}
\showDOI{%
\url{https://doi.org/10.1109/ICDE.2015.7113280}}


\bibitem[\protect\citeauthoryear{Solomonik, Besta, Vella, and
  Hoefler}{Solomonik et~al\mbox{.}}{2017}]%
        {solomonik2017scaling}
\bibfield{author}{\bibinfo{person}{Edgar Solomonik}, \bibinfo{person}{Maciej
  Besta}, \bibinfo{person}{Flavio Vella}, {and} \bibinfo{person}{Torsten
  Hoefler}.} \bibinfo{year}{2017}\natexlab{}.
\newblock \showarticletitle{Scaling betweenness centrality using
  communication-efficient sparse matrix multiplication}. In
  \bibinfo{booktitle}{{\em Proceedings of the International Conference for High
  Performance Computing, Networking, Storage and Analysis}}.
  \bibinfo{pages}{1--14}.
\newblock


\bibitem[\protect\citeauthoryear{Suzumura et~al\mbox{.}}{Suzumura
  et~al\mbox{.}}{2011}]%
        {suzumura2011performance}
\bibfield{author}{\bibinfo{person}{Toyotaro Suzumura} {et~al\mbox{.}}}
  \bibinfo{year}{2011}\natexlab{}.
\newblock \showarticletitle{{Performance characteristics of Graph500 on
  large-scale distributed environment}}. In \bibinfo{booktitle}{{\em Workload
  Char. (IISWC), IEEE Intl. Symp. on}}. \bibinfo{pages}{149--158}.
\newblock


\bibitem[\protect\citeauthoryear{Swamy et~al\mbox{.}}{Swamy
  et~al\mbox{.}}{2013}]%
        {swamy2013asymptotically}
\bibfield{author}{\bibinfo{person}{Vasuki~Narasimha Swamy} {et~al\mbox{.}}}
  \bibinfo{year}{2013}\natexlab{}.
\newblock \showarticletitle{{An Asymptotically Optimal Push--Pull Method for
  Multicasting Over a Random Network}}.
\newblock \bibinfo{journal}{{\em Inf. Theory, IEEE Tran. on\/}}
  \bibinfo{volume}{59}, \bibinfo{number}{8} (\bibinfo{year}{2013}),
  \bibinfo{pages}{5075--5087}.
\newblock


\bibitem[\protect\citeauthoryear{Tate, Kamil, Dubey, Gr{\"o}{\ss}linger,
  Chamberlain, Goglin, Edwards, Newburn, Padua, Unat, et~al\mbox{.}}{Tate
  et~al\mbox{.}}{2014}]%
        {tate2014programming}
\bibfield{author}{\bibinfo{person}{Adrian Tate}, \bibinfo{person}{Amir Kamil},
  \bibinfo{person}{Anshu Dubey}, \bibinfo{person}{Armin Gr{\"o}{\ss}linger},
  \bibinfo{person}{Brad Chamberlain}, \bibinfo{person}{Brice Goglin},
  \bibinfo{person}{Carter Edwards}, \bibinfo{person}{Chris~J Newburn},
  \bibinfo{person}{David Padua}, \bibinfo{person}{Didem Unat}, {et~al\mbox{.}}}
  \bibinfo{year}{2014}\natexlab{}.
\newblock \showarticletitle{Programming abstractions for data locality}.
\newblock  (\bibinfo{year}{2014}).
\newblock


\bibitem[\protect\citeauthoryear{Voss}{Voss}{}]%
        {tbb-pp}
\bibfield{author}{\bibinfo{person}{Michael Voss}.}
\newblock \bibinfo{title}{{Understanding the internals of tbb::graph :
  Balancing Push and Pull}}.
\newblock   (\bibinfo{year}{????}).
\newblock


\bibitem[\protect\citeauthoryear{Wang et~al\mbox{.}}{Wang
  et~al\mbox{.}}{2016}]%
        {wang2016hybrid}
\bibfield{author}{\bibinfo{person}{Zhigang Wang} {et~al\mbox{.}}}
  \bibinfo{year}{2016}\natexlab{}.
\newblock \showarticletitle{{Hybrid Pulling/Pushing for I/O-Efficient
  Distributed and Iterative Graph Computing}}. In \bibinfo{booktitle}{{\em ACM
  Intl. Conf. on Man. of Data}}. \bibinfo{pages}{479--494}.
\newblock


\bibitem[\protect\citeauthoryear{Whang et~al\mbox{.}}{Whang
  et~al\mbox{.}}{2015}]%
        {whang2015scalable}
\bibfield{author}{\bibinfo{person}{Joyce~Jiyoung Whang} {et~al\mbox{.}}}
  \bibinfo{year}{2015}\natexlab{}.
\newblock \showarticletitle{{Scalable Data-Driven PageRank: Algorithms, System
  Issues, and Lessons Learned}}.
\newblock In \bibinfo{booktitle}{{\em Euro-Par: Par. Proc.}}
  \bibinfo{pages}{438--450}.
\newblock


\bibitem[\protect\citeauthoryear{Yang and Leskovec}{Yang and Leskovec}{2015}]%
        {yang2015defining}
\bibfield{author}{\bibinfo{person}{Jaewon Yang} {and} \bibinfo{person}{Jure
  Leskovec}.} \bibinfo{year}{2015}\natexlab{}.
\newblock \showarticletitle{Defining and evaluating network communities based
  on ground-truth}.
\newblock \bibinfo{journal}{{\em Knowledge and Information Systems\/}}
  \bibinfo{volume}{42}, \bibinfo{number}{1} (\bibinfo{year}{2015}),
  \bibinfo{pages}{181--213}.
\newblock


\bibitem[\protect\citeauthoryear{Zaharia et~al\mbox{.}}{Zaharia
  et~al\mbox{.}}{2012}]%
        {Zaharia:2012:RDD:2228298.2228301}
\bibfield{author}{\bibinfo{person}{Matei Zaharia} {et~al\mbox{.}}}
  \bibinfo{year}{2012}\natexlab{}.
\newblock \showarticletitle{{Resilient Distributed Datasets: A Fault-tolerant
  Abstraction for In-memory Cluster Computing}}. In \bibinfo{booktitle}{{\em
  Proc. of the USENIX Conf. on Net. Sys. Design and Impl.}} {\em
  (\bibinfo{series}{NSDI'12})}. \bibinfo{pages}{2--2}.
\newblock
\showURL{%
\url{http://dl.acm.org/citation.cfm?id=2228298.2228301}}


\bibitem[\protect\citeauthoryear{Zhang et~al\mbox{.}}{Zhang
  et~al\mbox{.}}{2016}]%
        {zhang2016exploring}
\bibfield{author}{\bibinfo{person}{Mingxing Zhang} {et~al\mbox{.}}}
  \bibinfo{year}{2016}\natexlab{}.
\newblock \showarticletitle{Exploring the hidden dimension in graph
  processing}. In \bibinfo{booktitle}{{\em USENIX Symp. on Op. Sys. Des. and
  Impl. (OSDI 16)}}.
\newblock


\bibitem[\protect\citeauthoryear{Zhao}{Zhao}{2003}]%
        {zhao2003model}
\bibfield{author}{\bibinfo{person}{Yang Zhao}.}
  \bibinfo{year}{2003}\natexlab{}.
\newblock {\em \bibinfo{title}{A model of computation with push and pull
  processing}}.
\newblock \bibinfo{thesistype}{Ph.D. Dissertation}. \bibinfo{school}{Citeseer}.
\newblock


\bibitem[\protect\citeauthoryear{Zhu et~al\mbox{.}}{Zhu et~al\mbox{.}}{2016}]%
        {zhu2016gemini}
\bibfield{author}{\bibinfo{person}{Xiaowei Zhu} {et~al\mbox{.}}}
  \bibinfo{year}{2016}\natexlab{}.
\newblock \showarticletitle{{Gemini: A computation-centric distributed graph
  processing system}}. In \bibinfo{booktitle}{{\em USENIX Symp. on Op. Sys.
  Des. and Impl. (OSDI 16)}}.
\newblock


\end{thebibliography}

\maciej{One omp critical in SSSP is collapsed - remember this and make consistent for the ginal version}

\maciej{Conflict-Removal (CR)}
\maciej{with Nono}

\maciej{!!!!!!!!!!!!!!! for PageRank complexity: To get why there is this log
term, consider a fully connected bipartite G with |V1| much larger than |V2|}

\maciej{verify TC EREW }

\maciej{REMOVED:  Note that this categorization depends on how an algorithm is designed and not
on what it actually computes. For example, SSSP-$\Delta$ derives
distances from a source vertex. Yet, the former proceeds in iterations that
spans all the vertices; computing the shortest distance takes the form of
performing a simple operation on each vertex separately until it converges.
Contrarily, SSSP-$\Delta$ is similar to a traversal: in each
iteration, only the vertices from the current bucket are considered; new
vertices are processed once they are encountered by analyzing the edges of the
vertices from the current bucket.}

\maciej{Mention in general the LP simulation, Brent's Theorem, etc.}

\nono{Working on GHS. We have to note that GHS is particular implementation of
Boruvka/Sollin algorithm, moreover it is this one we can analyse in PP terms.
GHS algorithm uses message passing so quite obviously uses both push
(broadcast/send messages) and pull mechanisms, but it's important to note that
GHS not a "clear" example of defined PP meaning (it is not pure
"fine-grained"). In fact it is a specific kind of "bucket-driven" algorithm,
and it is always easier to find PP mechanisms in that case. As we said GHS
consists of two general types of operations. The first one is checking the
cheaper outgoing edge of a node; we probably could interpret that part as pull
mechanism: because we have to get (pull) values of branches of the node and
after that select the lightest one. The second one is to send (push) messages
with respectively requests and responses. Maybe there is a possibility to also
pull that messages but it seems to be more natural to simply broadcast them -
and so we have to decide is it useful to analyse that?}

\nono{There is another important thing. Note that generally MST algorithms are
difficult or impossible to analyse in the terms of PP because they are
"edge-oriented" (i.e. most common Kruskals' implementation bases only on edges
array and offers edges array as result). It is probably the main difference
between other MST algorithms and GHS, in which we use ordinary "node-oriented"
way. I suppose this difference is not only the result of quantity and method of
storage of the graph data, but also the way we are using to solve given
problem. But could we figure out for example BFS algorithm uses only edges?
I've tried to do it but I'm not sure if it is really possible. Mainly because
in this weird case edge is not object of operation but in some way this
operation itself (get parent node value (pull) - update child node (push)).}

\nono{Note that in some more complex algorithms - i.e. TC or CF - possibility of
using push and pull approaches sometimes depends on the method of
implementation. Moreover (and it is quite interesting), we can construct for
example TC algorithm completely suitable for our interpretation (in fact by make
it "fine-grained"). In case of TC there is an important phase of "passing"
adjacency tables between neighboring nodes. But maybe it is not necessary to
pass full table? Maybe it is better (and faster) to pull or push values from
these tables one by one from/to neighboring nodes? Actually this solution is
very similar to PR.}

\maciej{Algorithms are now more like work-span; they should be transformed into
``true'' barebone PRAMs}

\maciej{Add acknowledgements}

\maciej{we use PRAM for 'fine-grained' algorithms and BSP/LogP for messages}

\maciej{(is PRAM and BSP/LogP used for separate algs only? or is there some overlap?)}

\maciej{I think there is some fundamental difference between push and pull when
it comes to (1) pure atomic-issue (shared memory), and (2) messages (dist.
memory).  It seems that, for atomics, the difference between push/pull will be
the presence or absence of atomics; while for messages - the difference between
push/pull is the number of messages exchanged. We need to explore this as
well!! Some questions: (A) where/how locks come into this game? Meaning, we can
rewrite atomic-based schemes with locks. Does it change anything? (B) How does
it relate to asynchronous approaches/algorithms?}

\maciej{Is it more about fine-grained and coarse-grained push vs pull? fine
-atomics, coarse - messages?  How about RMA and locks here, and
shared/distributed memory distinction here??????}

\maciej{Buckets? 2) Push and pull can be really useful when we are using
algorithms that divide their vertexes into some buckets - as in mentioned SSSP
article. It is important to check it.}

\maciej{Memory tradeoff? In push PR MP, I can compute degree locally and send
the ready updates to the target. In Pull PR MP, I probably have to ask for
degree (slower) or keep all the degrees locally (more memory used)?}

\maciej{It seems that pull with messages in PR (MP) is useless? Twice the same
amount of communication?}

\maciej{In TC, I use an additional table of size n because of cache coherence problems (double check!)}

\maciej{Async PRAM?}

\maciej{Here, we need a general hybrid scheme (or schemes) that works for
any considered algorithm (or a class of algorithms) and switches between push and pull. Can we have it?}

\maciej{How does it apply to various types of graphs? Can we have some cool heuristics
that switches between these two? In what types of algorithms?}

\maciej{PRAM analysis: consider various schedules in Push in CREW?}

\maciej{Recheck the TC analysis!!! on SM!!! recompile}

\maciej{Can we take the inverse of A in the BFS in LA?} 

\maciej{In discussion more: Graph Distribution, Scope, Misconceptions}

\maciej{Geral PP?}

\maciej{\subsection{Push and Pull Misconceptions}}

\maciej{Add: where it does/doesn't make sense to switch between push and pull}

\maciej{Say how we fix all these statements in the literature.
One paper: this evaluation I read on the delta flight. Check also
these two surveys. Also, SSSP paper.}

\maciej{Correct Boruvka surname}

\maciej{The used representations (directed graphs) double memory, do we
change it?}

\maciej{\macbs{Utilized Functions}}

\maciej{\subsection{Label Setting and Correcting}}

\maciej{Check diameters of RMAT}

\maciej{Can we have different push/pull for communication AND synchronization?}
\maciej{CLAIM: we prove that the common view is wrong that some algorithms are inherently push
or pull based; cite some above.}

\end{document}